\documentclass[sigconf,nonacm]{acmart}
\usepackage{graphicx}
\usepackage{subcaption}
\usepackage{multirow}
\AtBeginDocument{%
  }

\setcopyright{acmlicensed}
\copyrightyear{2018}
\acmYear{2018}
\acmDOI{XXXXXXX.XXXXXXX}
\acmConference[Conference acronym 'XX]{Make sure to enter the correct
  conference title from your rights confirmation email}{June 03--05,
  2018}{Woodstock, NY}
\acmISBN{978-1-4503-XXXX-X/2018/06}

\begin{document}

\title{\textit{TaleBot}: A Tangible AI Companion to Support Children in Co-creative Storytelling for Resilience Cultivation}

\author{Yonglin Chen}
\orcid{0009-0004-3081-1813}
\affiliation{%
\institution{School of Design,}
  \institution{Southern University of Science and Technology}
  \city{Shenzhen}
  \country{China}
}
\email{12531641@mail.sustech.edu.cn}

\author{Jingjing Zhang}
\orcid{0009-0007-5405-9138}
\affiliation{%
    \institution{Picto AITech}
    \city{Hong Kong}
    \country{China}
}
\email{jingjing.zhang97213@gmail.com}

\author{Kezhuo Wang}
\orcid{}
\affiliation{%
\institution{School of Design,}
  \institution{Southern University of Science and Technology}
  \city{Shenzhen}
  \country{China}
}
\email{12333213@mail.sustech.edu.cn}

\author{Pengcheng An}
\orcid{}
\affiliation{%
\institution{School of Design,}
  \institution{Southern University of Science and Technology}
  \city{Shenzhen}
  \country{China}
}
\email{anpc@sustech.edu.cn}

\author{Xueliang Li}
\orcid{}
\authornote{Corresponding author.}
\affiliation{%
\institution{School of Design,}
  \institution{Southern University of Science and Technology}
  \city{Shenzhen}
  \country{China}
}
\email{lixl6@sustech.edu.cn}

% \affiliation{%

% \affiliation{%

% \affiliation{%

% \affiliation{%

% \affiliation{%

% \affiliation{%

% \affiliation{%

\renewcommand{\shortauthors}{Trovato et al.}

\begin{abstract}
Resilience is a key factor affecting children's mental wellbeing and future development. Yet, limited HCI research has explored how to help children build resilience through adversarial experiences. Informed by a formative study with elementary school teachers and professional psychologists, we design \textit{\textit{TaleBot}}, an AI-empowered system that supports children to co-create stories about overcoming everyday adversities tailored to their personal situations. We evaluated the system with 12 elementary children in school counseling rooms under teacher guidance and conducted reflective interviews with parents upon the Child–AI co-created stories. The findings show that \textit{TaleBot} encourages children in self-expression of feelings and thoughts, creating opportunities for teachers to provide personalized support and for parents to better understand the profound impact of family communication on children’s mental wellbeing. We conclude with design implications for using generative AI to support children’s mental health education and interventions across school and family contexts.
\end{abstract}



%   % \includegraphics[width=\textwidth]{sampleteaser}


\maketitle

% Introduction 的写作思路
% 1. 儿童心理健康的重要性 / 韧性教育的关键作用
% 2. AI赋能故事讲述在儿童心理健康教育中的应用 (叙事疗法)
% 3. 相关工作简要回顾
% 4. 提出研究问题

% 研究问题1 (系统如何运作?): 儿童和心理健康教育者如何参与AI支持的故事讲述？
% 数据收集方法: 调研、家庭配对访谈

% 研究问题2 (效果如何?): AI如何通过故事讲述在治疗环境中支持儿童韧性教育？

% 研究问题3 (设计启示): 如何在技术设计中整合AI，以联动儿童、教育者（治疗师）和家长共同推进韧性教育？

% 论文结构概述

\section{Introduction}

% 第一段：问题背景 - 心理健康对儿童成长至关重要，韧性教育的重要性
%Mental health for children is fundamental to their overall development and well-being \cite{who_unicef_mental_health_2024}. Resilience education has emerged as a critical approach to promoting children's mental well-being, with resilience theory demonstrating that all children can benefit from learning essential skills such as coping with stress and regulating emotions \cite{catalano2004positive}.

%Resilience refers to the ability of the individual to bounce back to original state of mind, or even better equiped with coping skills, after going through disadvantagous life situations [REF]. While early work resilience takes root in trauma-related resaerch on people living through childhood trauma [REF], recent research resilience takes insights from positive psychology, suggesting that resilience can be gained through daily adversarial experiences. 

Resilience has been reported to be a key factor indicating children's mental health and is highly relevant to their future life quality \cite{catalano2004positive, hanewald2011reviewing}. Children who demonstrate resilience are more inclined to develop a positive outlook in the face of challenges and to undergo growth after adversity, thereby enhancing their coping mechanisms and emotional maturity \cite{coyle2011resilient, hanewald2011reviewing, zolkoski2012resilience}. Despite its recognized importance, fostering resilience in children remains a longstanding challenge for parents and educators \cite{condly2006resilience, mallin2012mental}. Families play a crucial role in socializing children by equipping them with the skills to cope with stressors and recover from everyday adversities, such as small setbacks, performance pressure, and interpersonal conflicts, with parent–child communication serving as a key mechanism for cultivating children's appropriate responses to stressful circumstances and fostering resilience building \cite{sheridan2012understanding, theiss2018family}. The lack of positive parent-children interactions may undermine family wellbeing, leaving children more vulnerable to mental health challenges \cite{newland2014supportive}. On the other hand, traditional school-based approaches often rely on lecture-based teaching and post-adversity interventions, overlooking experiential learning that helps children build resilience as a life skill through self-reflection \cite{bhattachaya2013building, cahill2014building}. There is a need for more personalized, adaptive, and proactive school-based interventions to promote children's resilience.

%Resilient children are more likely to adopt a positive mindset towards challenging situations and experience post-adversity growth, emerging with stronger coping skills and emotional maturity 
%narrative therapy, expression therapy, and other methods that are constrained by physical materials and require intensive adult facilitation, making it challenging to deliver personalized and sustained psychological interventions.

% 第二段：故事讲述的潜力 - 故事讲述作为韧性教育的有效方法
Storytelling has been widely integrated in psychological interventions to assist individuals, especially children, to help them develop emotional regulation and problem-solving skills \cite{divinyi1995storytelling, mccall2019storytelling}. Through guided conversations with the professionals, storytelling provides safe and fictional contexts for children to develop narratives to make sense of difficult experiences, drive meaning from them, and build beliefs around these challenging situations \cite{ramamurthy2024impact, de2016digital}. However, traditional storytelling methods often require extensive one-on-one professional guidance, and are constrained by the availability of static materials, and limited in providing personalized content at scale \cite{ramamurthy2024impact}. 

%Existing digital mental health tools for storytelling were limited in combining the therapeutic and interactive features, making the experience both beneficial and engaging for the children [REF?].

% 第四段：研究机会 - 现有工作的局限性， 结合AI个性化能力与韧性教育原理的机会 ，以及和传统叙事疗法的区别 
%XL: I moved this paragraph here to make it more natrual and efficient for reading

%This gap presents a significant opportunity to leverage AI's adaptive capabilities for evidence-based psychological interventions. Traditional narrative therapy approaches, while effective, face several limitations: they require extensive one-on-one professional guidance, are constrained by static materials, and struggle to provide personalized content at scale [REF?]. Existing digital mental health tools often lack the interactive storytelling elements that make narratives engaging for children. By combining AI's personalization potential with established resilience education principles, we can create more effective and accessible tools that overcome these limitations, enabling dynamic, adaptive narratives that respond to individual children's needs while maintaining therapeutic rigor and supporting collaborative engagement among children, educators, and families.

% 第三段：AI技术的机遇与现状 - AI为故事讲述带来新机遇，但现有系统缺乏心理健康应用
Interactive storytelling systems have emerged as a significant intervention for promoting engagement of children in educational and creative activities in school and family contexts \cite{cai2023cathill, pordelan2023using}. Recent advances in artificial intelligence (AI) have opened new possibilities for enhancing such storytelling systems. AI-powered storytelling tools can provide personalized, adaptive narratives that respond to children's diverse needs and preferences \cite{10.1145/3613905.3650770}. These systems can enable collaborative story creation, supporting both creativity and purposeful narrative exploration \cite{10.1145/3706598.3713478}. Most existing AI-based storytelling systems in HCI focus on enhancing specific skills related to children's intellectual, social, and creative growth \cite{hagen2023evaluating}, such as critical thinking \cite{10.1145/3706598.3713602}, literacy \cite{10.1145/3613904.3642580}, creativity and expressive skills \cite{10.1145/3613904.3642852}. However, limited work exists that supports children's psychological resilience development. In addition, most of these systems only involve parents in peripheral roles without their perspectives and perceptions being fully investigated. More work is needed to understand how the child-AI interactions would engage parents in further reflection on their roles in family communication, in our case, for the purpose of building the children's resilience. There are following three research questions (RQs): 

%most of these systems are designed for one-on-one child-AI interaction, with minimal involvement of parents, who play a critical role in fostering children's resilience. In this study, we aim to fill this research gap through a design-based empirical study, driven by the following research questions:

% 第四段：研究问题 - 基于系统提出的核心研究问题
\begin{itemize}
    % RQ1: 探索性研究 - 儿童如何与AI支持的共同创作故事系统互动？
    \item \textbf{RQ1:} How do the children engage the AI-based interactive storytelling system for resilience education?
    % RQ2: Tale'Bot系统效果 - 心理健康教师和家长对使用生成式AI支持儿童韧性教育的看法如何？
    \item \textbf{RQ2:} How do the child-AI co-created stories engage parents in reflection on child-parent communication to better support children's resilience building?
    % RQ3: 设计启示 - 如何在技术设计中有效整合AI，以促进儿童、教育者和家长的协作式韧性教育？
    \item \textbf{RQ3:} What are the design implications of an AI-based interactive storytelling system for resilience cultivation?
\end{itemize}

% 第五段：系统介绍 - 介绍TaleBot系统的设计和功能
% 第七段：研究方法概述 - 研究方法和评估

We start our research with a formative study consisting of interviews with elementary children's parents, elementary school mental health teachers, mental health counselors, and psychiatrists. The results of this formative study provide a multi-stakeholder perspective on design requirements for an AI-powered interactive storytelling system for personalized resilience education. Based on these insights, we designed \textit{TaleBot}, a generative AI-powered system that engages elementary school children in collaborative storytelling with AI, driven by story outlines predefined by mental health teachers. \textit{TaleBot} features two interconnected platforms: a teacher-facing interface (backstage) that enables mental health teachers to customize story outlines for resilience education based on their assessment of each child’s personal circumstances, and a child-facing interface (frontstage) where children engage with a conversational AI to collaboratively generate a story book. During the child-AI interactions, the child is prompted by questions such as “What would you do in this situation if you were this character?”, and the response of the child serves as input for the generation of the story book unfolding on the interface chapter by chapter. Unlike existing storytelling systems focused solely on software, we gave our design a physical form with a 3D-printed PLA shell and a furry fabric outer layer, housing the tablet with its screen facing outside (Fig. \ref{fig:study_env_overview}a). This transforms our design into an embodied conversational agent that children can hold and touch during interactions.

%Additionally, AI and psychological counselors collaborate to produce mental health reports based on children's responses during storytelling interactions and their personal circumstances, which are provided to parents along with feedback collection to better understand each child's psychological development and provide actionable recommendations.

With \textit{TaleBot} as the technology probe, we conducted an empirical study involving one school teacher (who is also the school counselor), 12 elementary children, and 5 parents of the children. Our prototype was deployed to the counseling room in an elementary school as an assistive tool for the teacher to help the elementary children deal with various everyday adversities. 12 children (6 male, 6 female; aged 7 to 8) were invited to interact with  \textit{TaleBot} to co-create stories with the story outlines predefined by the teacher prior to the session. Observation of the co-creative sessions and post-session interviews with the teacher provided insights into how the system engaged children in expressing personal feelings, thoughts, and coping strategies through imaginary yet personally relevant situations, creating further opportunities for teacher guidance and education (RQ1). After each session, the co-created story printed out into a story book for the child to take home as a probe for child-parent communication. In addition, an digital version of the story book annotated with teacher and AI's commentaries was shared with the parent separately. Based on this, semi-structured interviews were conducted with the parents to understand how the child-AI co-created stories engaged the parents in reflection on their parental practices better support children’s resilience development (RQ2). Discussions upon the findings of the study provide implications for future design of AI-empowered interactive systems for supporting children's resilience cultivation (RQ3).

%Findings of the study revealed that the children... \textcolor{red}{[to be filled in after data analysis]} (RQ1). From the study, we also gathered perspectives from mental health teachers and parents on using the system's generative AI approach for resilience education (RQ2). 

% 第八段：伦理声明 - 研究伦理审批和数据保护措施
%This study received IRB approval from university, with multi-stakeholder consent protocols involving parents, teachers, psychiatrists, and clinical psychologists. All identifiable data were anonymized through pseudonym replacement and secure storage.

% 第九段：主要贡献
This work makes three key contributions to the HCI community: (1) \textbf{System design and implementation} of \textit{TaleBot}, an AI-powered interactive storytelling system to engage children in collaborative storytelling with an converataional agent to create personalized stories about overcoming daily adversities, thereby fostering resilience development; (2) \textbf{Empirical insights} on how children engage with such AI-powered interactive storytelling systems, and multi-stakeholder perspectives of the teachers and parents towards the implementation of such designs resilience education; and (3) \textbf{Design implications} for future AI-powered mental health interventions that provide personalized, adaptive experiences for children, with appropriate involvement of multiple stakeholders across school and family contexts.
\section{Related Work}

\subsection{Resilience Building through Everyday Adversities for Children}

Psychological resilience, also referred to as mental resilience or individual resilience, refers to the ability of the individual to bounce back to the original state of mind, and even better equipped with coping skills, after going through disadvantageous life situations \cite{graber2022psychological, cooper2013building, ijntema2019reviewing, fletcher2013psychological}. In particular, Bryan et al. \cite{bryan2019stressing} describe resilience as “a dynamic process encompassing the capacity to maintain regular functioning through diverse challenges or to rebound through the use of facilitative resources.” While early work resilience takes root in trauma-related research on people living through traumatic experiences, such as natural disasters and childhood abuse \cite{levine2009examining, tedeschi1996posttraumatic}, recent research on resilience takes insights from positive psychology, suggesting that resilience can be gained through everyday adversities, such as setbacks and work-related stressors \cite{arslan2024embracing, fletcher2013psychological, yu2019personal}. Literature on resilience building emphasizes interventions before, during, and after adversity, which is a "necessary condition" for resilience \cite{luthar2000construct, southwick2011interventions}, and argues for a mindset that "embraces change and accepts that it is a possible outcome of adversity" \cite{ijntema2019reviewing, ho2025shine}. This is especially the case for children who demonstrate great psychological flexibility while going through challenging situations, if provided with appropriate support and guidance. Coping with and gaining self-growth from these adversarial situations compose a significant part of the trajectory of children's psychological development, enabling them to acquire essential skills to cope with various challenges in future life \cite{catalano2004positive, condly2006resilience, dubowitz2016adversity}. 

Resilience cultivation is a significant component in children's mental healthcare in schools \cite{doll1998risk, brown2021resilience, fenwick2018systematic}. However, traditional school-based resilience-building programs for children focus on didactic instruction and post-crisis interventions \cite{fenwick2018systematic}. Such approaches overlook the experiential nature of resilience, which is best promoted through reflection on real-life experiences, and may leave the children at risk of long-term effects of unsolved mental health challenges. This indicates design opportunities for HCI technologies for providing more situated, personalized, and adaptive solutions for promoting children's mental health \cite{children5070098}.

\subsection{Digital and AI-empowered Mental Health for Children in HCI}

%Children's mental health challenges, including depression, anxiety, bullying, and social-emotional competence deficits, have become increasingly prevalent in recent years. Traditional therapeutic approaches often face barriers such as limited accessibility, high costs, and stigma, creating a significant treatment gap in child mental healthcare \cite{sit2024digital}. To address these challenges, 

The HCI community has explored a variety of digital technologies to improve children's mental health. These applications focus on a range of barriers faced by children, such as a lack of mental health literacy, limited parental care, constrained access to professional services, and social stigma \cite{sit2024digital}. On the one hand, we have seen overarching, expert-led systems such as screening tools for early detection of mental health issues \cite{10.1145/3613904.3642604}, and VR applications that allow children to enter carefully designed simulated scenarios where they can practice coping with distressing situations in safe, controlled environments \cite{halldorsson2021annual}. On the other hand, there are child-centered technologies that support children to gain mental health knowledge and psychological capabilities through child-parent interactions \cite{slovak2018just} and serious games with therapeutic skills integrated in the gameplay \cite{SCHONEVELD2016321}. While these digital mental health tools proved their potential to benefit children in their original contexts, they often lack the adaptive and personalization capabilities needed to address the diverse and evolving needs of individual children.

With recent advances in Generative AI technologies, such as Large Language Models (LLMs) and text-to-image generation models, AI-based systems have shown the advantage of providing more personalized and appropriate mental health support to children \cite{dray2024review, wanniarachchi2025personalization}. These systems can engage children in natural, conversational interactions and creative activities that are tailored to their individual needs and preferences, while aligned with therapeutic goals. For example, Hui et al. \cite{10.1145/3628516.3659399} present an AI-empowered conversational agent to measure and foster children's physiological resilience through guided conversations integrating personal and social support factors. \textit{DiSandbox} \cite{10.1145/3706598.3713660} is an AI-powered sandbox system that supports children in sandbox play under the guidance of AI, while their creations can be used for mental health assessment and to engage parents for timely interventions. \textit{EmoEden} \cite{10.1145/3613904.3642899} is an AI-based training system to support children with high-functioning autism in emotional learning and social skill development through personalized conversations. 
%Other examples include...

Despite the increasing number of AI-empowered systems for supporting children's mental health, it seems that most applications focus on guided conversations using LLMs, focusing on risk prediction or screening while following an expert-led approach \cite{dray2024review, mansoor2025conversational}. The need for providing personalized, adaptive, and proactive interventions is not fully met by these applications. And, to our best knowledge, only a few (\cite{10.1145/3628516.3659399}) place resilience building as the central effort in designing the systems. We extend our review to digital storytelling systems, which show great potential for delivering personalized and engaging experiences for children while fostering resilience by revisiting adversity and cultivating a positive mindset about such experiences in safe, controlled environments.

\subsection{AI-empowered Storytelling Systems}

%While AI-child interactions have shown promise in individual therapeutic contexts, children's mental health support often requires a broader ecosystem involving multiple stakeholders who each bring unique perspectives, needs, and capabilities to the therapeutic process. Storytelling, as a fundamental human activity for meaning-making and emotional expression, has emerged as a particularly powerful medium for multi-stakeholder collaboration in children's mental health interventions. The integration of AI into storytelling processes creates new opportunities for coordinated support that leverages the strengths of different stakeholders while addressing their individual limitations.

%From the therapeutic professional perspective, AI serves as a powerful tool for enhancing expressive arts therapy and facilitating multi-sensory therapeutic interactions. Research demonstrates that AI can function as embodied material for expressive arts therapy, promoting multi-sensory interaction and assisting family communication while strengthening collaboration and empathy between all participants \cite{10.1145/3613904.3642852}. This approach illustrates how AI can augment therapists' capabilities by providing dynamic, responsive tools that support therapeutic storytelling processes involving multiple family members, creating richer and more engaging therapeutic experiences than traditional individual therapy sessions.

Interactive storytelling systems, also referred to as storyreading systems by some researchers \cite{chen2025characterizing}, are platforms designed to support users in co-creation or interactive reading of stories, often embedded with educational purposes with children as target users. Based on how these systems involve users in interaction with the key story elements, Fan et al. \cite{10.1145/3706598.3713478} characterize these systems into three types: emergent storytelling systems where users influence the story development by selecting from predefined narrative elements (characters and backgrounds), character-driven storytelling systems where users enact the stories as one of the characters, and user-centered plotting systems where users dynamically shape he story narratives through decision-making as the stories unfold.

AI-empowered storytelling systems show greater potential to provide personalized and engaging experiences for children in interaction with such systems. These applications can be understood by examining the different roles played by AI agents in assisting children in co-creating and reading stories. In particular, we have seen how AI assists children in the visualization of stories, as a story illustrator or a visual storyteller, through the integration of text-to-image or image-to-image models. Like \textit{StoryDrawer} \cite{zhang2022storydrawer}, which engages the children in further elaboration of story elements and transforms these into drawings or sketches. These systems show the advantage of motivating children in imagining and elaborating ideas, while the interactions may rely on children's verbal or visual expression abilities \cite{10.1145/3706598.3713478}. In addition, there are systems where AI, as a plot-writing assistant, supports children in the co-construction of story narratives. For example, \textit{StoryPrompt} \cite{fan2024storyprompt, 10.1145/3706598.3713478} enables the children to set up characters and settings, and to shape the storyline through keywords suggested by the system or their own real-time inputs. Another example is \textit{Mathemyths} \cite{10.1145/3491102.3517479}, which is a voice-based conversational agent that engages children in story co-creation through open-ended questions and scaffolding responses, while embedding mathematical knowledge into the unfolding narrative. Challenges in designing such systems are to ensure the consistency and flow of the storylines while incorporating children's inputs, which can be random or outside the intended scope. In addition to the above systems that involve the children in the co-creation of story elements, there are also systems that take a more peripheral role, engaging children in conversations around the stories as reading partners. For example, \textit{StoryMate} \cite{chen2025characterizing} engages the children in guided conversations, connecting the story context to real-world knowledge to support children's active thinking. Other similar examples are \textit{StoryBuddy}\cite{zhang2022storybuddy}, and \cite{hu2024grow}.

%Mutimodal storytelling with physical prompts, such as sketching [StoryDrawer] and clay [Liu et al., 2024]. involvement of adult stakeholders, Parents or teachers work as converstational guide in collaborative storytelling, by asking prompting questions about the story and provide responses to the outcomes of the creation [StoryBuddy]. 

These works demonstrate the potential of storytelling systems to provide personalized reading experiences and foster meaningful conversations that support a wide range of skills. However, to the best of our knowledge, little research has explored their application in the context of resilience education except \cite{hu2024grow} where the AI agent is only involved as a reading partner. In addition, most systems follow a child-centered approach with a lack of consideration for the involvement of multi-stakeholders. While some studies include parent participants, such as parents and teachers (e.g., \cite{chen2025characterizing}), their roles are mostly peripheral, limited to roles as reading companions or instructors. Little is known about their perspectives on how such designs could be integrated into family contexts or their own roles in supporting children’s development, in this case, the cultivation of resilience.

\section{Formative Study}

As a preparation for the design phase, we conducted a formative study consisting of semi-structured interviews with key stakeholders in children's resilience education. The study served two purposes: first, to understand the current landscape of children's adversity experiences and existing support mechanisms; second, to inform the design of digital storytelling platforms that support involvement of multi-stakeholders for supporting the children's resilience education.

\subsection{Method}

We recruited seven stakeholders: five mental health professionals and two elementary school teachers. The professionals included one child psychiatrist (26 years of experience), three psychological counselors from an independent agency (each with 10+ years of experience in child/adolescent counseling), and one school-based counselor. Both teachers had 25+ years of teaching experience and dealt with students' mental health issues as homeroom teachers. All participants had substantial experience with children's mental health. Semi-structured interviews lasted 30-60 minutes. Participants answered four core questions based on their experiences: (1) common adversity scenarios faced by children, (2) maladaptive and adaptive coping responses, (3) current prevention and intervention methods, and (4) perspectives on digital platforms in resilience education. All interviews were audio-recorded with the participants' consent. The interview data were analyzed using systematic thematic analysis methods \cite{hsieh2005three}. We followed an iterative process of coding, categorizing, and theme development to identify key insights, which inform design requirements for digital systems aimed at supporting children’s resilience.

% 压缩前版本We recruited a total of seven stakeholders, including five mental health professionals and two elementary school teachers. These professionals include 1 child psychiatrist (with 26 years of working experience), 3 psychological counselors from an independent counseling agency (each with over 10 years of experience in child and adolescent counseling), and 1 school-based counselor. The two elementary school teachers have over 25 years of teaching experience (one is an English teacher and the other is a Mathematics teacher), and both have experience dealing with students' mental health issues as homeroom teachers over the years. All participants had substantial experience working on dealing with children's mental health issues. The semi-structured interviews with each participant lasted 30-60 minutes. During the interviews, participants were asked to these four core questions based on their first-hand experiences with children: (1) common adversity scenarios faced by children, (2) maladaptive and adaptive coping responses manifested by children, (3) current prevention and intervention methods for children's mental health, and (4) their perspectives on the use of digital platforms in resilience education. All interviews were audio-recorded with the participants' consent. The interview data were analyzed using systematic thematic analysis methods \cite{hsieh2005three}. We followed an iterative process of coding, categorizing, and theme development to identify key insights, which inform design requirements for digital systems aimed at supporting children’s resilience.

\subsection{Key Findings}

%\subsubsection{Sources of Psychological Challenges}
%Children face adversities across multiple domains. In family contexts, conflicts arise from parental control over daily schedules and screen time, with parents often imposing their own standards on children without considering the child's needs and preferences. Family system disruptions, such as parental divorce, further compound these challenges. School-based conflicts include teacher-student tensions, such as being criticized, ignored, or wrongfully accused. As one counselor noted: \textit{"A third-grade student's deskmate brought playing cards to school and left them in his drawer. When the teacher found them, she criticized him, but he said it wasn't his."} Peer conflicts, including bullying and academic competition, also contribute to children's stress, often exacerbated by limited psychological knowledge among both children and their peers.

\subsubsection{
    % \textcolor{red}{rephrase by XL} 
KF1: Children face diverse family and school adversities.} Our interviews revealed several adversities contributing to children's psychological challenges. These include family conflicts from parents imposing strict rules on schedules and smartphone use. Many children seeking psychosocial support came from "problematic" families where parents ignore or inappropriately handle their children's emotional issues. Children also face adverse school experiences like peer relationship tensions or teacher conflicts. One counselor described a child who became depressed after being falsely accused by a teacher of bringing playing cards to class, with parents uncertain how to help.

% \subsubsection{ 压缩前版本
%     % \textcolor{red}{rephrase by XL} 
% KF1: Children are challenged by various family and school adversities} From our interviews, we identified several adversities that may contribute to children’s psychological challenges. These include family conflicts arising from parents imposing strict rules on their children’s schedules and smartphone use. According to the psychological counselors, many children who came for psychosocial support were from "problematic" families, where parents either ignore or handle their children’s emotional issues inappropriately. We have also noticed that children might be affected by adverse experiences at school, such as tensions in peer relationships or conflicts with teachers. One counselor recalled a child who became depressed after being falsely accused by a teacher of bringing playing cards to the classroom, while the parents were unsure how to resolve the situation.

%\subsubsection{Children's Coping Responses}
%Children exhibit diverse behavioral responses to adversity. Avoidance behaviors include school refusal and withdrawal from communication with adults. Aggressive responses manifest as physical fights with peers or, when emotions remain unprocessed, self-harm behaviors. Many children also display heightened fear and sensitivity. However, children with better psychological education demonstrate help-seeking behaviors as a constructive coping strategy.

\subsubsection{
% \textcolor{red}{Rephrase by XL} 
KF2: Appropriate Support to help children elaborate their feelings and thoughts.} Many participants emphasized the importance of supporting children in expressing their own thoughts and perspectives before drawing conclusions about their situations. The psychiatrist, counselors and teachers shared their techniques in communicating with the children. For example, \textit{"To understand from a child's perspective, what exactly is it that makes them uncomfortable."} Counselors also use the method of scene reproduction:\textit{"Use the sand table method or the painting method to present the child's anger."}

\subsubsection{KF3: The Dual Nature of Adversity and Support for Post-adversarial Growth}
All participants agreed that adversity serves as both challenges and opportunities for children's long-term self-growth. The psychiatrist noted: \textit{"Such setbacks shape children's personality formation and may better prepare them for future life."} A counselor compared resilience building to muscle training: \textit{"Like building muscle, overcoming difficult situations with positive feedback promotes growth in stress tolerance and cognitive thinking."} However, this counselor emphasized that excessive negative feedback is detrimental: \textit{"Continuous negative experiences are like constantly looking in a mirror reflecting 'I am not good enough.' This repeated negative reinforcement limits children's potential. While moderate failure teaches valuable lessons, success breeds success - positive experiences create encouraging feedback that transfers to future challenges, making children progressively more capable."} The counselor stressed that adversity should be accompanied by adequate support systems to ensure growth rather than trauma.

% 压缩前\subsubsection{KF3: The Dual Nature of Adversity and Support for Post-adversarial Growth}
% All the participants agreed that adversity serves as both challenges and opportunities for the children's self-growth in the long term. The psychiatrist noted: \textit{"Such setbacks, whether positive or negative, shape the children's personality formation and may prepare them better for their future life."} A counselor compared resilience building to muscle training: \textit{"Like building muscle, when people overcome difficult situations with positive feedback, it promotes growth in stress tolerance and cognitive thinking."} However, this counselor also emphasized that excessive negative feedback can be detrimental: \textit{"Continuous negative experiences are like looking in a mirror that constantly reflects 'I am not good enough, I am incapable.' This repeated negative reinforcement limits children's potential. While moderate failure can teach valuable lessons, success breeds success - positive experiences create pleasant emotions and encouraging feedback that transfer to future challenges, making children progressively more capable."} The counselor stressed the importance of moderation: adversity should be accompanied by adequate support systems to ensure growth rather than trauma.

\subsubsection{KF4: Barriers and Limitations in Current Support Systems}

We recognize each role's limitations in supporting children through adversities. One counselor noted it was "too late" when parents "sent" children to them, expecting professional intervention to simply "fix" issues rooted in unhealthy family communication. Teachers mentioned heavy teaching loads prevented addressing students' underlying psychological needs, while child-parent conflicts were often the source of depression. The psychiatrist emphasized changing entrenched parental behaviors was the biggest obstacle: \textit{"The most difficult part is changing parents' fixed cognition... I advise them not to communicate with criticism... But they unconsciously revert back to blaming the child... Sometimes even causing secondary harm."} Teachers could only provide momentary emotional comfort without follow-up. All participants noted lacking communication among teachers, counselors and parents toward shared understanding and collaborative efforts. \textit{"Because some parents have particularly bad tempers... as soon as children come home from school, their parents' faces tend to be gloomy, which can easily lead to family conflicts... some unsafe incidents may occur... As teachers, apart from providing psychological counseling to students, we also offer methodological guidance to parents."} These limitations underscore the need for alternative, child-centered approaches complementing adult-mediated interventions.

%\subsubsection{Design Requirements for Digital Platforms}
%Participants provided specific recommendations for resilience education platforms, emphasizing an emotion-first approach. As one counselor emphasized: \textit{"Start with emotions first. Children need to be seen, listened to, and understood. When children's identity systems conflict with school and family expectations, they won't feel seen or understood."} This emotional validation must be embedded in relatable, familiar contexts that mirror children's actual experiences.

%Beyond emotional support, participants stressed the importance of engagement and intrinsic motivation. Platforms must be sufficiently engaging and enjoyable to encourage children's voluntary participation and sustained interaction. To achieve this, one counselor specifically recommended role-playing mechanisms: \textit{"Let children step into different roles and recreate scenarios, similar to sandplay therapy."} This approach allows children to safely explore and process their experiences through narrative and imaginative play.

%Finally, platforms should offer guided, scaffolded experiences with highly personalized difficulty levels that incorporate everyday scenarios, ensuring that each child's unique needs and developmental stage are appropriately addressed.

\subsubsection{KF5: Child-centered Design with Open-ended Interfaces for Adults' Participation}

All participants agreed on a child-centered approach for designing digital systems supporting children's mental wellbeing and resilience. One counselor emphasized: \textit{"Start with emotions first. Children need to be seen, listened to, and understood. When children's self-identity conflicts with school environment or family expectations, they may feel overlooked or misunderstood."} This emotional suppression must be addressed in contexts where children develop self-understanding and growth through active learning. Participants emphasized encouraging voluntary participation and self-exploration through engaging interactions. One counselor recommended incorporating role-play elements, allowing \textit{"children to step into different roles in simulated scenarios, similar to sandplay therapy."} Open-ended systems should provide personalized experiences responding to children's unique circumstances and psychological needs while enabling scaffolded support, as \textit{"some children think it's just a minor conflict... However, some children need a week of continuous psychological support."}

% 压缩前版本All the participants agreed on a child-centered approach towards designing digital systems to support children's mental wellbeing and resilience building. As one counselor emphasized: \textit{"Start with [understanding children's] emotions first. Children need to be seen, listened to, and understood. When children's sense of self-identity conflicts with the school's environment or family expectations, they may feel overlooked or misunderstood."} This suppression of emotional expression and personality must be addressed in the context where they may develop deeper self-understanding and thus self-growth through active learning. The participants also emphasized the significance of encouraging voluntary participation and self-exploration through engaging and enjoyable interactions. Regarding this, another counselor recommended incorporating role-play elements into the design, allowing "the children to step into different roles in simulated scenarios, similar to sandplay therapy." On the other hand, every child has a different ability to feel adversity, open-ended systems that can provide personalized experiences for children (in response to their unique circumstances and psychological needs), while enabling them to provide scaffolded support. like \textit{"Some children think it's just a minor conflict and that everything will be fine once it's resolved. However, some children need a week of continuous psychological support. "}

\subsection{Design Requirements}
These insights directly inform five design requirements (DR) for developing personalized AI-empowered storytelling platforms for supporting children's resilience building: \textbf{DR\#1:} personalization of story structures and development that suit individual children's circumstances and psychological needs (in response to KF1, KF4 and KF5); \textbf{DR\#2:} engaging and enjoyable experiences for children through interactive storytelling and child-friendly narratives and visual elements(KF5); \textbf{DR\#3:} role-playing mechanism for emotional engagement and elaboration of personal feelings and thoughts (KF2, KF4); \textbf{DR\#4:} multi-stakeholder platforms to promote conversations and shared understanding across different groups(KF4, KF5).

%These insights directly inform five design requirements (DR) for developing personalized AI-empowered storytelling platforms for supporting children’s resilience building: DR#1: personalized storyline setting and calibration (addressing the variety of school and family diversity, adversity’s dual nature and open-ended interfaces for adults’ participation); DR#2: engaging experiences through interactive storytelling (child-centered design); DR#3: appealing visual design with relatable characters (child-centered design); DR#4: personally relevant story contexts reflecting children’s real experiences (addressing barriers and limitations in current support systems and child-centered design); DR#5: role-playing mechanisms for safe exploration (child-centered design).
\section{Designing \textit{TaleBot}}

\subsection{System Design}

Based on the five design requirements (DR) derived from our formative study, we developed \textit{TaleBot}, an interactive storytelling system that engages children in co-storytelling with AI. To fulfill \textbf{DR\#1}, the system features an expert-facing interface, serving as the backstage of the child-AI interaction, which allows the teacher (the school counselor) to customize the story outlines according to each child’s unique circumstances and psychological needs. For \textbf{DR\#2}, \textit{TaleBot} provides a child-facing interface that combines children’s story book narratives, visual elements, and a child-friendly conversational AI. Interaction with the system enables children to co-create a story book with AI, which unfolds in more detail according to the children's responses to TaleBot's prompting questions at key story milestones. We also give the child-facing interface a physical form that strengthens the social presence of the conversational AI (Figure 1. left). In line with \textbf{DR\#3}, interaction with the system may encourage the children to project themselves into the story contexts, thus fostering self-expression of feelings and thoughts under similar circumstances. As for \textbf{DR\#4}, the process and outcomes of the child-AI story co-creation can provide opportunities for child-teacher communication during the co-creative process and  child-parent conversations on the storybooks after the interactions. Below, we elaborate on specific design features with more details and the underlying rationale. 

\begin{figure*}[h]
    \centering
    \includegraphics[width=1\textwidth]{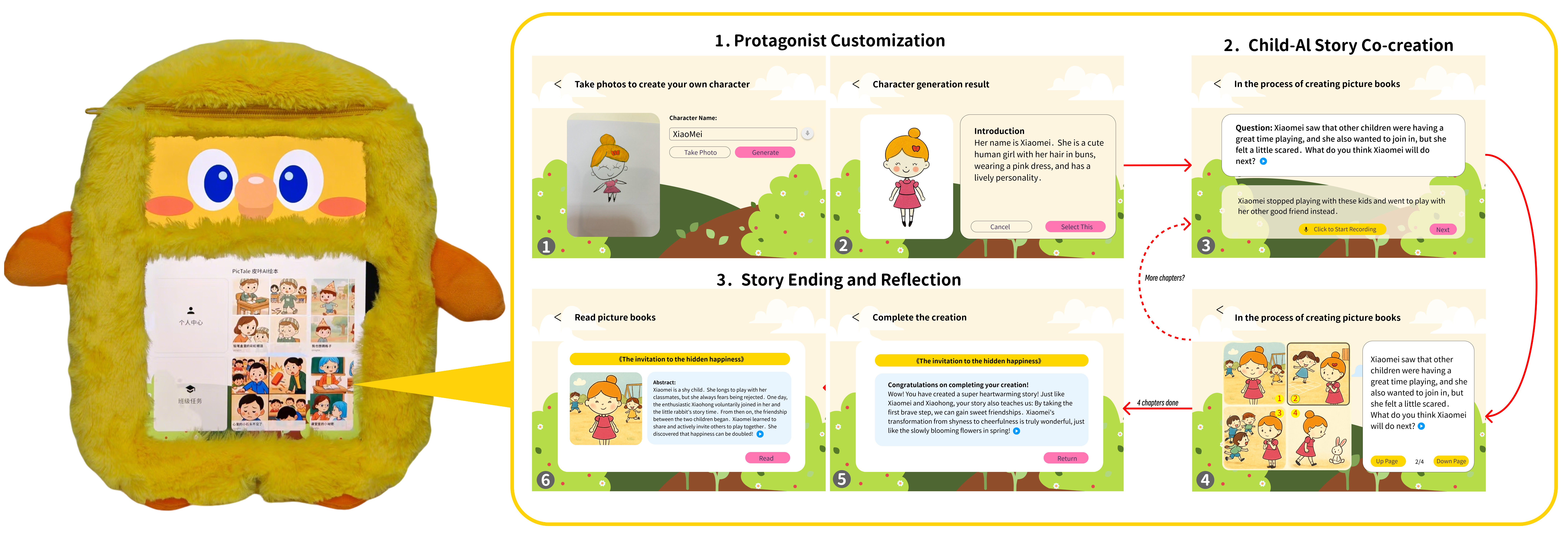}
    \caption{The TaleBot interface (left) and three primary interaction modes (right).}
    \label{fig:talebot_interface}
\end{figure*}

%\subsubsection{Mental Health Teacher as the Plot Setter}

%suggestion for the struture of this section:

%1.Multi-faceted interfaces for teacher-guided personalized storytelling (overview of the architectural design and interaction process; structure and workflow of multiple agents)

%2. Branching Narrative Construction in the Pre-Storytelling Stage(expert-facing interface)

%3.A Voice-based, LLM-powered chatbot to support personalized input (LLM promoting for taking in child user interface)

%4.Chapter-based generation for adaptive reflection (prompt design for visualization generation)

%5.Tangible AI companion for enhanced social presence

% 统一叫法：
%  专家端 教师端：expert-facing interface.
%  儿童端？用户端？学生端？ child-facing interface.
%  物理形式 psysical forms

\subsubsection{Interface Design}
%\subsubsection{Multi-faceted interfaces for teacher-guided personalized storytelling}
% (overview of the architectural design and interaction process; structure and workflow of multiple agents)

\textit{TaleBot} consists of two primary interfaces: an expert-facing interface for the psychology teacher and a child-facing interface for interactive storytelling. The child-facing interface is implemented as a tablet-based application with a Flutter frontend and Python Flask backend architecture. For our user study, the software runs on iPads equipped with custom-designed enclosures featuring plush materials, as shown in the Figure \ref{fig:talebot_interface}.

\begin{figure*}[h]
     \centering
     \includegraphics[width=1\textwidth]{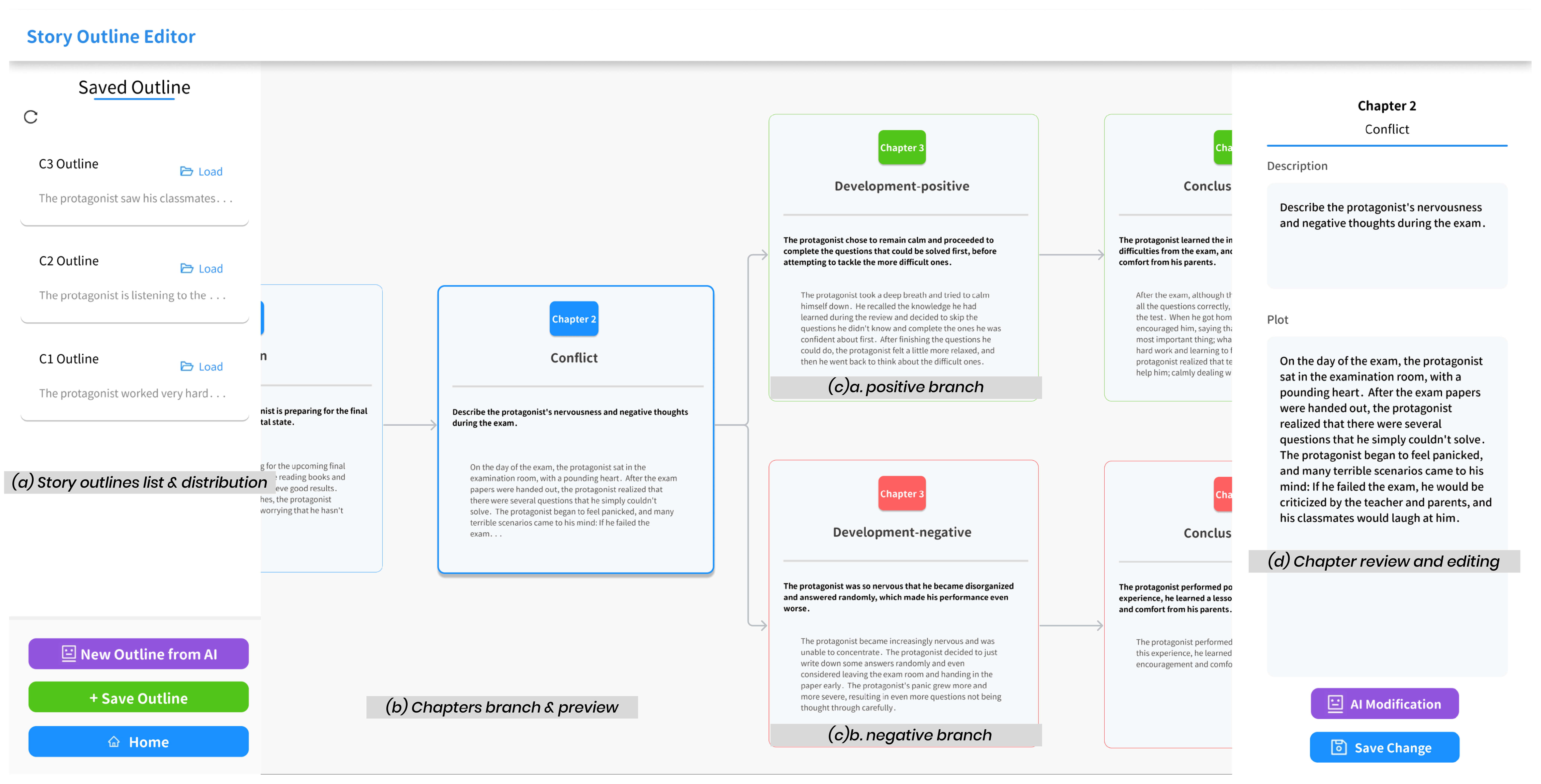}
     \caption{The expert-facing interface allows expert to customize personalized storylines for child-AI interactions. Panel (a) shows the outline list that can load data and distribute content to the child-facing interface. Panel (b) displays chapter previews and branching options. Panel (c) shows a story outline customized for participant C1, where Chapter 3 contains two branches: (c)a leads to a positive storyline, while (c)b leads to a negative storyline. Panel (d) demonstrates that when users click on a chapter in the preview panel, they can edit setting descriptions and plot details, or use AI assistance for modifications.}
     \label{fig:study_outline}
 \end{figure*}

The expert-facing interface is a React-based web platform that facilitates story outline generation and task deployment. This interface allows experts to generate four-chapter story frameworks through simple text input that describes the setting and plot of the story, serving as the foundation of the story co-creation between the child and TaleBot. Figure \ref{fig:study_outline} shows the interface design for the experts (teacher) to customize a story outline based on analysis of a child's situation and psychological character.

%For example, when a child is experiencing high academic pressure and perfectionist tendencies due to strict parenting, the expert (mental health teacher) might input a simple description: \textit{"A child[protagonist] who studies very hard for final exams but encounters many difficult problems during the test."} This input generates the structured therapeutic narrative shown in Figure \ref{fig:study_outline}. Once saved and published in the expert-facing interface, the story framework is immediately sent to the child-facing interface.

\subsubsection{AI Architecture Design}

%\subsubsection{A LLM-powered and Voice-based chatbot to support personalized input}

\begin{figure*}[h]
    \centering
    \includegraphics[width=1\textwidth]{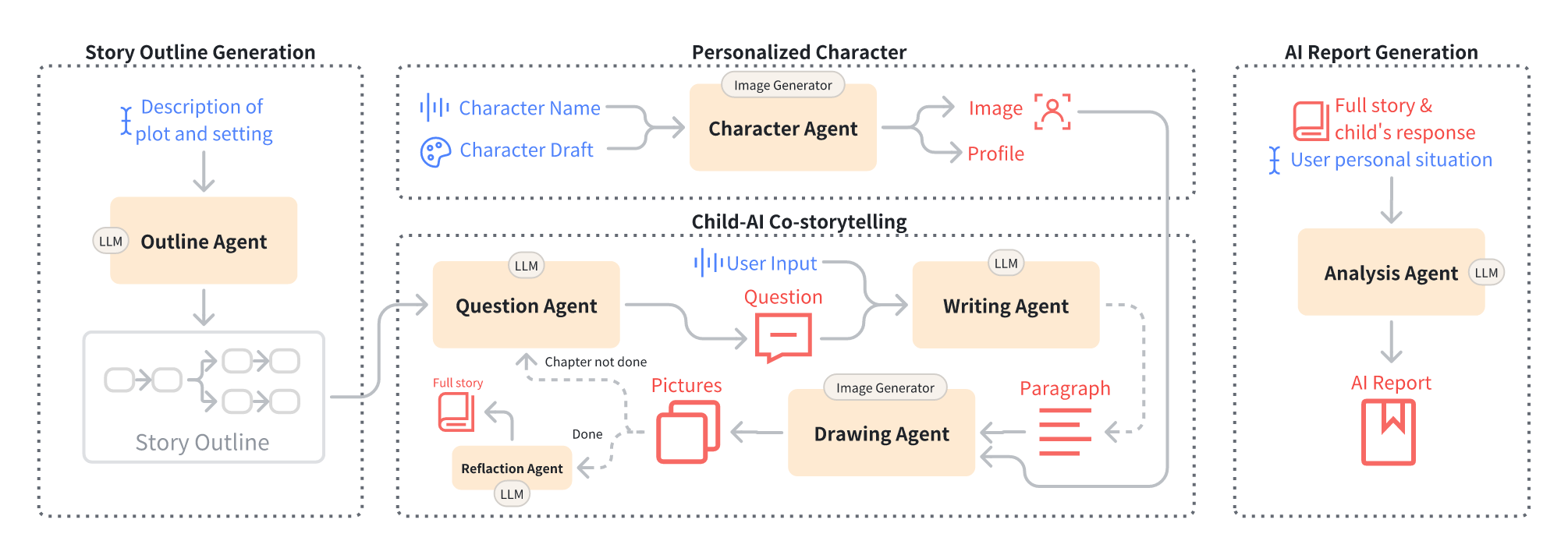}
    \caption{The inputs, outputs, and workflows of seven agents.}
    \label{fig:agent_framework}
\end{figure*}

%At the underlying logic of LLM-powered workflows, we designed several specialized agents to respond to AI storytelling system requirements. 

We designed a multiple-agent system integrating seven agents to support the generation of the stories with inputs from different sources. Figure \ref{fig:agent_framework} illustrates the workflow of this system. Detailed prompt specifications for each agent are provided in Appendix \ref{app:prompt-templates}.

\begin{itemize}
    \item \textbf{Outline Agent (Expert-facing)}: This agent generates story outlines which extends the teacher's input (brief description of a story) into a more structured storyline featuring four chapters. This structured story outline drives the development of the story with its details enriched by the children's responses to TaleBot's prompting questions at key story milestones.

%The input comprises concise plot and setting descriptions provided by experts specifying core narrative requirements. These inputs feed into the \textit{Outline Agent}, which function is to generate narrative structures like initiation, conflict, development, and conclusion. The generated outline annotates each chapter section with paragraph attributes and writing guidelines, provided for reference by subsequent child-facing agents.

    \item \textbf{Character Agent}: The agent transforms the children's drawings into story book-style illustrations of characters with a description generated from analysis of the drawing. The image and profile (description) of the character are constantly referred to by the other agents to ensure the consistency of image and text generation.

    \item \textbf{Question Agent}: This agent generates corresponding prompting questions depending on the story outlines and character profiles and the current progress of the story development.

    \item \textbf{Writing Agent}: This agent synthesizes the children's responses, the story outlines and character profiles into fully elaborated stories, which are represented by means of four paragraphs corresponding to the story scenarios.
    % \item 
%multiple inputs—the questions from the \textit{Question Agent}, transcribed user responses, the current chapter outline, and character profiles—to generate coherent story text structured in four paragraphs.
    \item \textbf{Drawing Agent}: Based on inputs of Writing Agent and the character image, this agent generates story scenarios in forms of four images with a four-panel layout.
    \item \textbf{Reflection Agent}: This agent generates the summary of the story with compliments to the child for finishing the story generation, while highlighting the educational meanings of the story.
    \item \textbf{Analysis Agent}: This agent generates a psychological analysis of the children's responses throughout the story co-creation, and provides advice to parents on how to better support the children to overcome similar situations in the future.

%Based on the complete story, the child's personal situation, and the child's responses for each chapter as input to the Analysis Agent, it will subsequently output a psychological analysis report that interprets potential psychological conditions and provides recommendations for parents.
\end{itemize}

% The models and prompts used by these agents during the experimental phase are detailed in the Appendix.

All questions and story narration within the system are delivered through a child-friendly storytelling voice, integrating the Doubao-voice-synthesis API for text-to-speech conversion. Compared to existing storytelling systems that use predefined or real-time generated keywords as narrative prompts (e.g., \cite{10.1145/3706598.3713478}), we utilize voice input to capture children's natural language during Child-AI conversations. This functionality also reduces reading requirements, particularly beneficial for young children with limited literacy skills.

Regarding the selection of large models, \textit{TaleBot} selected Deepseek V3 for text generation based on the consideration of its strong performance in Chinese language, while OpenAI's GPT-Image-1 was chosen for image generation to ensure consistent character portrayal and high visual quality.

%\subsubsection{Chapter-based generation for adaptive reflection}

%The storytelling process begins with prompt-driven large language model generation based on established story book characters and story frameworks. In the first phase of each chapter, the system generates guiding questions designed to encourage children to create chapter-specific narratives while exploring their coping strategies and emotional responses to adversity situations.

At the key milestones of the story development (prior to the generation of each chapter), \textit{TaleBot} poses a prompting question, aksing what the child would do if he or she were in the same situation, The child's response, captured through Tencent ASR API for accurate speech recognition, is then integrated as the input for the story generation (as mentioned above). Below is an example of Child-AI interaction:

\begin{quote}
\textbf{TaleBot:} \textit{Bunny is working hard to review for final exams, but as the exam date approaches, it begins to feel nervous and anxious. What do you think Bunny will do next?}

\textbf{Child user:} \textit{Take deep breaths. (C1)}
\end{quote}

% The system employs Tencent ASR API for accurate speech recognition. For generated questions, here is an example where a child created a character called "Bunny," and the question posed for the first chapter of the story shown in Figure \ref{fig:study_outline} is:

%This creates a closed loop of adversity scenario questioning and seeking children's solutions.

%After the child's response, the system enters the second phase, generating chapter content in real-time based on the story framework, posed questions, and the child's answers. After generating chapter text, the system creates four illustrations for that chapter, with character appearances consistently reflecting the child's original character design. This iterative process spans four chapters, enabling children to create a complete story book with a coherent narrative arc. Upon completion, the system presents a summary page providing an overview of the created story and offers positive reinforcement for both the story book content and the child's creative contributions, providing relatively positive affirmation of the child's solutions.

\subsubsection{Tangible AI companion for enhanced social presence}

%Additionally, existing tablets present several pain points: first, their form factor as a flat device does not represent an image that children easily approach or find appealing. Moreover, positioning them on a table for convenient reading requires additional stands. Inspired by plush toys, we considered how to design the device to resemble a stuffed animal. 

To give a physical form to the AI-powered system, we designed a tablet case (with a slot open on its top) taking the appearance of a cute animal, covered with the skin made of soft and fuzzy fabrics (as shown in the Figure \ref{fig:talebot_interface} left.). With this physical form, the child can interact with TaleBot by placing it on the table or holding it in the arms while feeling the soothing experiences touching its fuzzy skin.

The screen is split into two sections: the upper part displays eye and mouth animations, and the lower part serves as the operation area. During AI processing delays, the upper section shows a thinking animations, while during voice playback, it displays speaking animations.

%To achieve this effect, we first modeled an inner shell that supports tablet insertion, facilitating child-friendly angles for interaction and reading. Subsequently, 

%To better integrate hardware and software, we designed the plush exterior to resemble a chick, while the software implements a vertical screen division (1:1.6 ratio), as shown in the Figure \ref{fig:talebot_interface} left. The upper portion displays dynamic eyes that animate through software to show different states: speaking (during text-to-speech), listening and thinking (during user voice input), and idle states. The lower portion serves as the core interaction area.

\subsection{Design Evaluation}

\subsubsection{Participants}

%We recruited 12 child participants (ages 6-8, grades 1-2) from an elementary school, all of whom were receiving psychological counseling support. Participants were selected based on their current emotional and behavioral challenges, ensuring diverse representation of common childhood adversities identified in our formative study.

We recruited 1 mental health education teacher (the school counselor), 12 elementary students, and 5 of their parents through collaboration with a public elementary school. The school serves around one thousand students covering grades of 1 to 6, and is equipped with two mental health education teachers, who are responsible for delivering mental health literacy curricula across 6 grades and providing counseling services to students. One of the mental health education teachers was involved in the study as the instructor to assist the children in interaction with the system, as well as the psychological counselor to analyze the children's psychological needs and provide support during the child-AI interactions. The students were recruited through two channels: (1) the students who came to the counseling center seeking counseling services, and (2) those recommended by their homeroom teachers, who had identified potential psychological needs. After an initial talk with the children to understand their situations and confirming their interest in participation in the study, the teacher then contacted their parents to obtain parental consent. An ongoing recruitment strategy was adopted over a period of 3 weeks. A total of 12 elementary students  (6 Male, 6 Female; Mean = 7.333 years old, SD = 0.492) and 5 parents (including 1 parent who was a sibling, C10 and C11) voluntarily participated in the study with the consent of their parents. The teacher is well acquainted with all the child participants to ensure a comfortable and trustworthy atmosphere in the study. Table \ref{tab:participants_info} shows the demographic information of these participants. The study procedure was approved by the ethics committee of the first author's affiliated university.

\begin{table*}[t]
\centering
\caption{Participants demographic information}
\begin{tabular}{lccc|cccc}
\toprule
\textbf{Child ID} & \textbf{Gender} & \textbf{Grade} & \textbf{Age} & \textbf{Parent ID} & \textbf{Gender}& \textbf{Age} & \textbf{Occupation}\\
\hline
C1 & F & Grade 2 & 8 & P1 & M & 38 & Finance \\
\hline
C2 & F & Grade 2 & 8 & --- & --- & --- & ---\\
\hline
C3 & M & Grade 1 & 7 & --- & --- & --- & ---\\
\hline
C4 & M & Grade 1 & 8 & P4 & F & 48 & Housewife\\
\hline
C5 & M & Grade 1 & 7 & --- & --- & --- & ---\\
\hline
C6 & F & Grade 1 & 7 & --- & --- & --- & ---\\
\hline
C7 & M & Grade 1 & 7 & --- & --- & --- & ---\\
\hline
C8 & M & Grade 1 & 8 & P8 &M  & 39 & Marketing\\
\hline
C9 & F & Grade 1 & 7 & --- & --- & --- & ---\\
\hline
C10 & F & Grade 1 & 7 & \multirow{2}{*}{P10} & \multirow{2}{*}{M} & \multirow{2}{*}{39} & \multirow{2}{*}{AI Product Manager}\\
\cline{1-4} 
C11 & F & Grade 1 & 7 & & & &\\
\hline
C12 & M & Grade 1 & 7 & P12 & M & 35 & Marketing\\
\bottomrule
\label{tab:participants_info}
\end{tabular}
\end{table*}

\subsubsection{Study context}
\begin{figure*}[h]
    \centering
    \begin{subfigure}{0.58\textwidth}
        \centering
        \includegraphics[width=\textwidth]{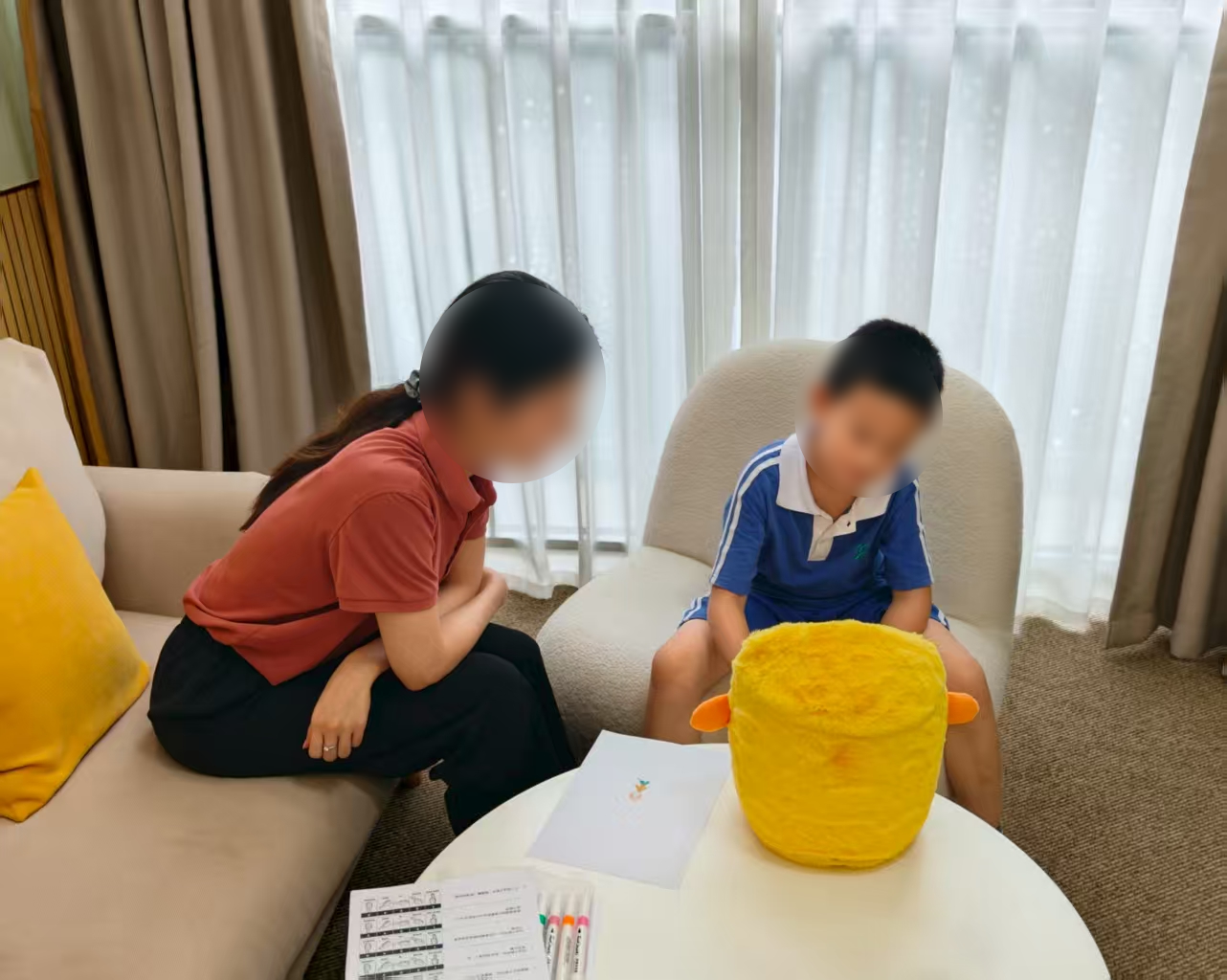}
        \caption{}
        % \caption{The teacher was guiding the child in interaction with \textit{TaleBot}.}
        \label{fig:study_env}
    \end{subfigure}
    \hfill
    \begin{subfigure}{0.35\textwidth}
        \centering
        \includegraphics[width=\textwidth]{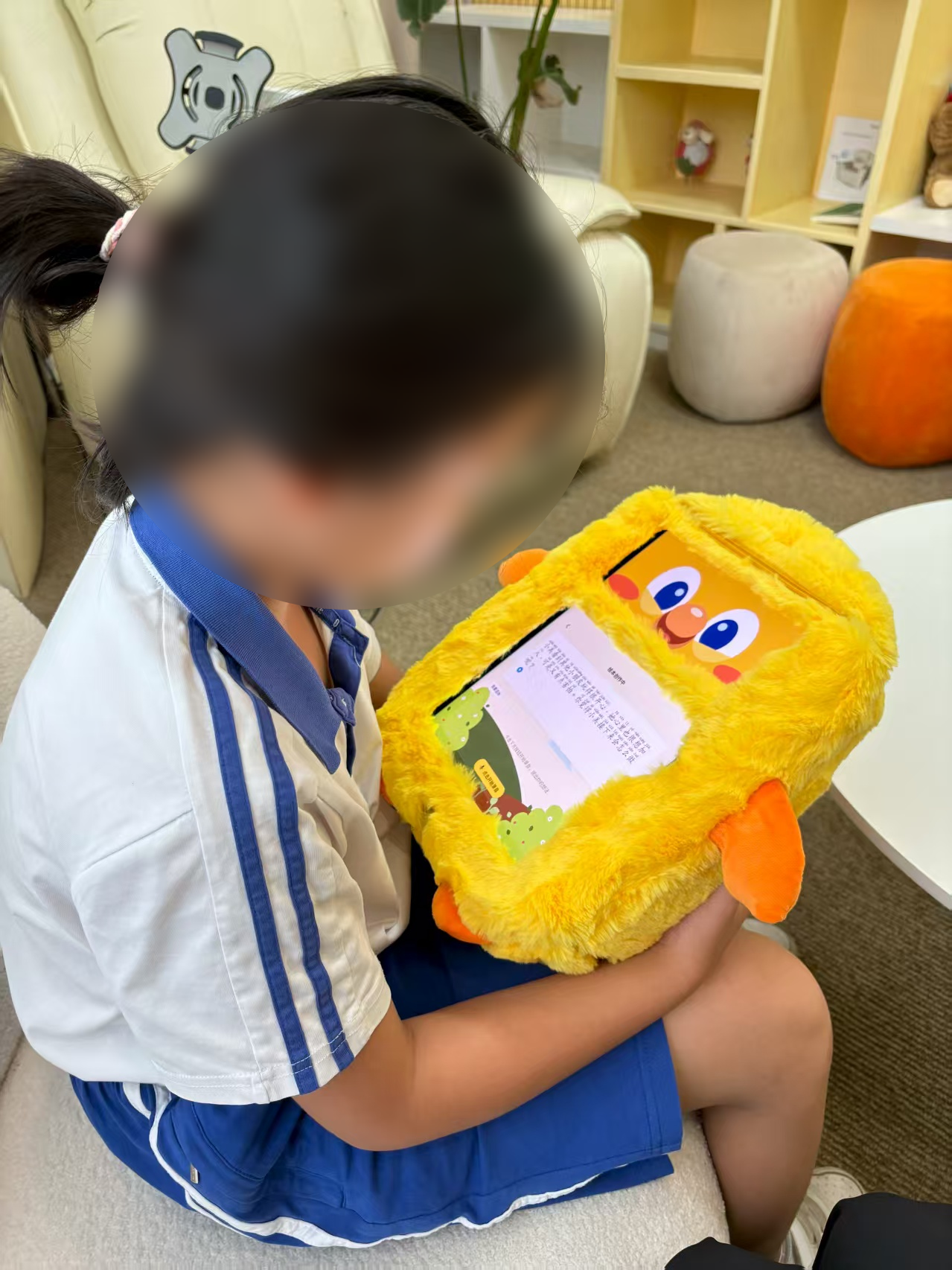}
        \caption{}
        % \caption{The child held the \textit{TaleBot} in his arms.}
        \label{fig:study_hold_example}
    \end{subfigure}
    \caption{The figure illustrates the study setup in a school counseling room: (a) the teacher (school counselor) (left) guided the children (right) in co-creating a story with \textit{TaleBot} placed on the table of the counseling room, and (b) the child interacted with \textit{TaleBot} in the arms.}
    \label{fig:study_env_overview}
\end{figure*}

The user study was conducted in one of the counseling rooms of the mental health center of this primary school. The room was equipped with sofas, tables, and a warm and comfortable environment. While the psychological teacher provided guidance nearby, like Figure \ref{fig:study_env_overview}a,  and the \textit{TaleBot} children could be placed on the table or held in the arms, like Figure \ref{fig:study_env_overview}b. 

% [Please add: what kinds of mental health service provided by the mental health center? and how does it work for the children to access the mental health service?]

% Figure * shows the school counseling room where the user study was conducted.

% [Figure *. Photo of the counseling room where the user study is conducted]

\subsubsection{Study Procedure}

\begin{figure*}[h]
    \centering
    \includegraphics[width=1\textwidth]{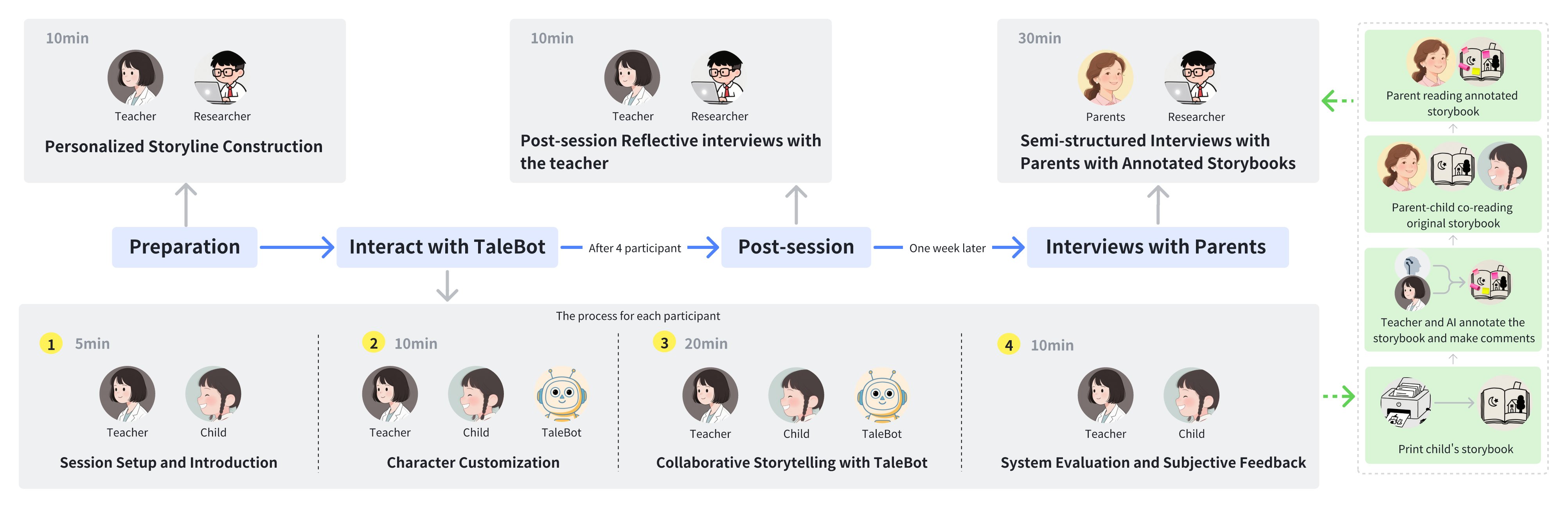}
    \caption{Study procedure.}
    \label{fig:study_procedure}
\end{figure*}

Prior to the study, the responsible researcher (first author of this paper) had several rounds of preparation meetings with the mental health teacher to ensure that she understood the study procedure and became familiar with the \textit{TaleBot} system. During the study, the teacher took two roles: as an instructor, guiding the children in their interaction with the system, and as a psychological counselor, engaging in conversations with them to understand their feelings and thoughts while offering appropriate advice. Figure \ref{fig:study_procedure} illustrates the overview of the procedure, highlighting the involvement of the teacher, the children, and their parents at different steps of the study.

%storyline customization, Child-AI co-storytelling, storybook production for child-parent communication, and interviews with the parents based on annotated storybooks.

%Each study session followed the same standardized procedure. To ensure children's comfort and engagement, psychological counselors who were familiar with and trusted by the children led the sessions, while researchers remained in an adjacent room on standby. 

%When children arrived at the experimental room, they participated in the study within the counseling room under the guidance of the psychological counselor.

\textbf{Personalized Storyline Construction:} Before each session, the teacher established a profile for each child participant through analyzing the child's personality, current psychological situations, and recent adversities they have experienced (Table \ref{tab:story_overview}). If needed, the teacher communicated with the homeroom teacher to better understand the children's current circumstances. With the support of the researcher, the teacher then transformed these personal profiles into story outlines using the expert platform. The teacher was encouraged to create multiple branches in the later stage of the story development, depending on the children's responses to key story milestones (as indicated in Figure \ref{fig:study_outline}).

%If the child was familiar to the psychological counselor, they could directly describe the child's specific situation. If the psychological counselor was less familiar with the child, they would write the personal profile with assistance from the child's homeroom teacher.

\textbf{Session Setup and Introduction:} When children arrived at the school counseling room, the researcher activated the recording equipments, including iPad screen recording, a GoPro camera, and an audio recorder to document the session. After the setup, the researcher left the room and remained on standby next door to provide technical support when needed, ensuring the child feels comfortable with only the teacher present. As the beginning of each session, the teacher introduced \textit{TaleBot} to the child with the following statement: \textit{"Today you will experience a toy that lets you tell stories with AI! The AI will ask you four questions as the story continues, and you can answer based on your own ideas. At the end, the AI will create a story book based on your answers."} After this, the teacher demonstrated how to interact with \textit{TaleBot} through co-creating a one-chapter story in front together with the child. Once the child indicated readiness, the formal study began.

%The teacher then demonstrated how to co-create stories with \textit{TaleBot} through, using one chapter as a case study to explain the four modules: AI questioning, user response, text generation, and image generation. Once the child indicated readiness, the formal experiment began.

\textbf{Character Customization:} The co-creation with \textit{TaleBot} began with character customization, in which a visual of the protagonist was generated from the child’s watercolor drawing. The teacher introduced this step by saying: \textit{"Now we need a protagonist for the story. You can use these watercolor pens to draw what you want the protagonist to look like."} During this process, the teacher prompted the children to think of their favorite animals or cartoon characters as the inspiration, and to assign a name to the protagonist, while ensuring the drawings were completed within 20 minutes. Next to this, the teacher assisted the child in to upload the drawing to \textit{TaleBot} which converted it into a more visually polished character, serving as the main character of the story to be co-created. This process could be repeated multiple times within the allocated time until the child is satisfied with the outcome.

\textbf{Collaborative Storytelling with \textit{TaleBot}:} \textit{TaleBot} engaged the children in collaborative storytelling by prompting questions at key story milestones, such as \textit{"what would you do if you were [protagonist's name] in this situation?"}. Based on the inputs of the children, the system generated a story chapter, consisting of four images. During this process, the teacher provided appropriate guidance to ensure children's attention remained focused on creation and addressed any difficulties they encountered in the interaction. The co-creation of stories ended after four chapters had been generated.

\textbf{System Evaluation and Subjective Feedback:} After the children finished the story creation, the teacher guided the children to provide feedback on their interaction experiences through the System Usability Scale (SUS) for Children \cite{putnam2020adaptation} and open-ended questions. To accommodate the children’s difficulty in understanding the scale and responding with numerical values, the teacher administered the questionnaire through verbal and gestural communication (as indicated in Appendix \ref{app:sus-scale}). The open-ended questions focus on the children's general experiences interacting with the system, their likes and dislikes during the process, their perceptions of the co-created stories, and how they would like to share these stories with others. The session ended with the teacher complimenting the children for their participation, ensuring they left the counseling room in a positive emotional state. 

\textbf{Post-session Reflective Interviews with the Teacher:} After the child exited the counseling room, the responsible researcher engaged the teacher in a reflective interview. The questions focused on: 1) the teacher's general impression of children's behavior during the session; 2) challenges and difficulties encountered in guiding the children in interacting with the system; 4) insights into the personality, coping behavior and psychological needs of the children as revealed through the session; and 3) the teacher's perception of how the sytem has facilitated the child-teacher communication on relevant topics. The reflective interviews provided a supplementary view on our analysis of children's behavior during the interaction sessions, and revealed a teacher's perspective on our design.

%\textbf{Session Conclusion:} Each session ended with researchers thanking participants: \textit{"Our experience today is now complete, thank you for participating. We will print your created storybook and deliver it to you later."} All recording equipment was then deactivated, including screen recording, GoPro camera, and audio recorder.

\textbf{Semi-structured Interviews with Parents with Annotated Storybooks as Conversational Probes:} After the interaction session, each child received a printed storybook adapted from their co-created stories (with adjustments to the layout of visuals and fonts), and carried it with them to share with their parents. In addition to this, the researcher prepared an annotated version of the storybook for the parents  (Appendix \ref{app:ai_analysis_report}), incorporating the child’s real-time responses at key story milestones, the teacher’s commentary on these responses, an analysis of the child’s overall situation, and the advice provided to the parents. Alongside the teacher’s annotations, we also included AI-generated comments to provide a comparative perspective on the same content. The annotated storybooks were delivered digitally to ensure exclusive access for the parents. 
% See Appendix X for an example of these annotated storybooks. 

%Additionally, based on the story books produced by children and their responses during the interaction process, we had psychological counselors provide commentary, analyzing the children's possible psychological states. At the end of the report, both AI and psychological counselors offered suggestions for parents. These psychological reports were then delivered to parents through the psychological counselors. Subsequently, we recruited parents willing to participate in interviews.

A semi-structured interview was conducted with the parents one week after they received the annotated storybooks. All the interviews were conducted via online calls. The core interview questions include: 1) What is your impression of the storybook? 2) What is your response to the teacher and AI's interpretation of the child-AI interaction and their physiological situations? 3) How is your approach to family communication to promote your child's mental health and cultivate their resilience? 4) What is your attitude towards AI-supported mental health interventions? 

\subsubsection{Data Analysis}

We collected various types of data from the study, including the original response text in the children's interaction, completed questionnaires, audio recordings of the child–teacher interactions, and semi-structured interviews with parents. We employed a mixed-methods approach to formulate an understanding of how the system engaged the children in child-AI co-creative interactions to cultivate resilience (RQ1) and the parents' attitudes and perspectives towards the AI-based interactive storytelling system for resilience cultivation (RQ2).

\textbf{Quantitative Analysis:} The SUS for Children responses were analyzed using descriptive statistics, calculating means and standard deviations across all participants. We conducted normality testing using the Shapiro-Wilk test to validate our statistical approach. The resulting SUS scores were interpreted according to established usability benchmarks \cite{bangor2008empirical} to classify the system's overall usability performance.

% <<<<<<< HEAD
% \textbf{Qualitative Analysis:} Our qualitative data consisted of interview transcripts and observational notes collected during the study sessions. We applied a systematic thematic analysis approach following established qualitative research methodologies [cite]. The analysis process involved phases: Two researchers, both native Chinese speakers fluent in English, independently reviewed all textual data to develop familiarity with the content. During this phase, we conducted open coding by labeling significant quotations with descriptive keywords and phrases in English. Firstly, The two researchers collaboratively organized the initial codes into coherent themes using affinity mapping techniques [cite]. This process involved revisiting original quotations and researcher memos to identify overlaps, resolve conflicts, and address disagreements. Redundant codes were removed while new insights emerging from collaborative discussions were incorporated into the coding framework. And then, Using the refined coding scheme developed in the previous phase, both researchers systematically processed all transcripts, matching quotations to appropriate codes and themes. This structured approach ensured comprehensive coverage of the data while maintaining consistency in interpretation.Finally, The coding process yielded a total of [X] open codes derived from [X] quotations across all participant families and volunteers, which were organized into [X] overarching themes. These themes form the foundation for our qualitative findings presented in the results section.
% =======

\textbf{Qualitative Analysis:} We applied the systematic thematic analysis approach \cite{10.1145/3544548.3581203} to analyze the transcripts of the audio recordings and observation notes. The analysis process involved four iterative phases. First, two researchers (the first and second authors of this paper), both native Chinese speakers fluent in English, independently reviewed all textual data to develop familiarity with the content. During this phase, we conducted open coding by labeling significant quotations with descriptive keywords and phrases in English. Second, the two researchers collaboratively organized the initial codes into coherent themes using affinity mapping techniques \cite{10.1145/3322276.3322326}. This process involved revisiting original quotations and researcher memos to identify overlaps, resolve conflicts, and address disagreements. Redundant codes were removed while new insights emerging from collaborative discussions were incorporated into the coding framework. Third, a third and fourth researcher (the fourth and corresponding author of this paper) joined the discussion meetings to refine the coding scheme developed in the previous step, to ensure a shared understanding within the research team. Following this, the two researchers systematically processed all transcripts, matching quotations to appropriate codes and themes. This structured approach ensured comprehensive coverage of the data while maintaining consistency in interpretation. As a result, the data analysis yielded a total of 164 open codes derived from transcripts of the original recordings across all participating children, parents, and teacher, which were organized into 15 overarching themes. These themes form the foundation for our qualitative findings presented in the results section.
% >>>>>>> b3a20298f8da513972444ab14c7cb346ec6365e9

\section{Results}

The findings of this study are presented in two parts in response to the two research questions. In Section 5.1, we demonstrated how \textit{TaleBot} engaged the children in the co-creation of stories, which encouraged them to express their feelings and coping strategies in relevant situations (\textbf{\textit{RQ1}}). We also incorporate the teacher's perspective towards the system and their experiences of supporting the children in the interaction. In Section 5.2, we elaborated on the parents' reflection on the Child-AI co-created stories and their views on the potential role of AI-empowered systems for resilience education in family contexts (\textbf{\textit{RQ2}}).

\subsection{Children's engagement and experiences with the system (\textbf{\textit{RQ1}})}

\subsubsection{An overview of Child-AI co-created stories}
Based on previous research literature \cite{gunning2000creating}, we look into the stories in terms of the three key story elements in this research, i.e., \textbf{settings}, \textbf{plots}, and \textbf{characters}. The settings and plots were predefined by the teacher based on their understanding of the children’s personal circumstances and recent daily adversities, and the characters were generated from the children’s drawings and prompt inputs. Appendix \ref{app:story_overview} shows an overview of the settings, plots, and characters of the stories predefined by the teacher, and co-created by the children and \textit{TaleBot}. Overall, the teacher customized 12 story outlines for all the children, combining their personal circumstances, personalities, and recently experienced adversities. The co-created narratives unfold across familiar environments, with \textbf{settings} in school predominating: classrooms (C2, C4, C5, C10), playgrounds (C3, C7, C9, C11, C12), examination rooms (C1), and teacher offices (C6, C8). Two stories (C7, C10) extend into home environments, bridging school and family contexts.

The \textbf{plots} of these stories explore four interconnected dimensions of childhood adversity, with individual narratives often addressing multiple challenges simultaneously. Academic performance anxiety drives plots about test-related stress and fear of failure (C1, C4, C5), while emotional dysregulation through scenarios involving anger, frustration, and mood management (C2, C3, C6, C8, C9, C11). Peer interaction challenges shape plots about friendship and social inclusion (C3, C7, C9, C10, C11, C12), and behavioral compliance issues create plot tensions around conflicts with authority and rule-following (C6, C8, C12). This plot mirrors children's lived experiences, where academic, emotional, social, and behavioral challenges rarely occur in isolation.

The \textbf{characters} created based on children's preferences demonstrate diversity. While most children (n=10) chose human characters as their story avatars—comprising six boys (C3, C4, C5, C7, C8, C12) and four girls (C2, C6, C9, C11)—two children opted for animal protagonists: a rabbit named Bunny (C1) and a kangaroo called Xiaoai (C10). All protagonists were designed to mirror the children's own developmental stage, facilitating identification and engagement.

%Children first drew their envisioned characters by hand, which were then digitally recreated using the AI system. The character selection process resulted in diverse protagonists: animal characters appeared in 2 stories, including a cute rabbit (C1) and a little kangaroo (C10), while human characters dominated with 10 stories featuring age-appropriate protagonists. The gender distribution showed 8 male protagonists (C3, C4, C5, C6, C7, C8, C12) and 4 female protagonists (C2, C9, C11).

\subsubsection{Children's interaction experiences with the system.} 
\begin{figure*}[htbp]
\centering
\includegraphics[width=1\textwidth]{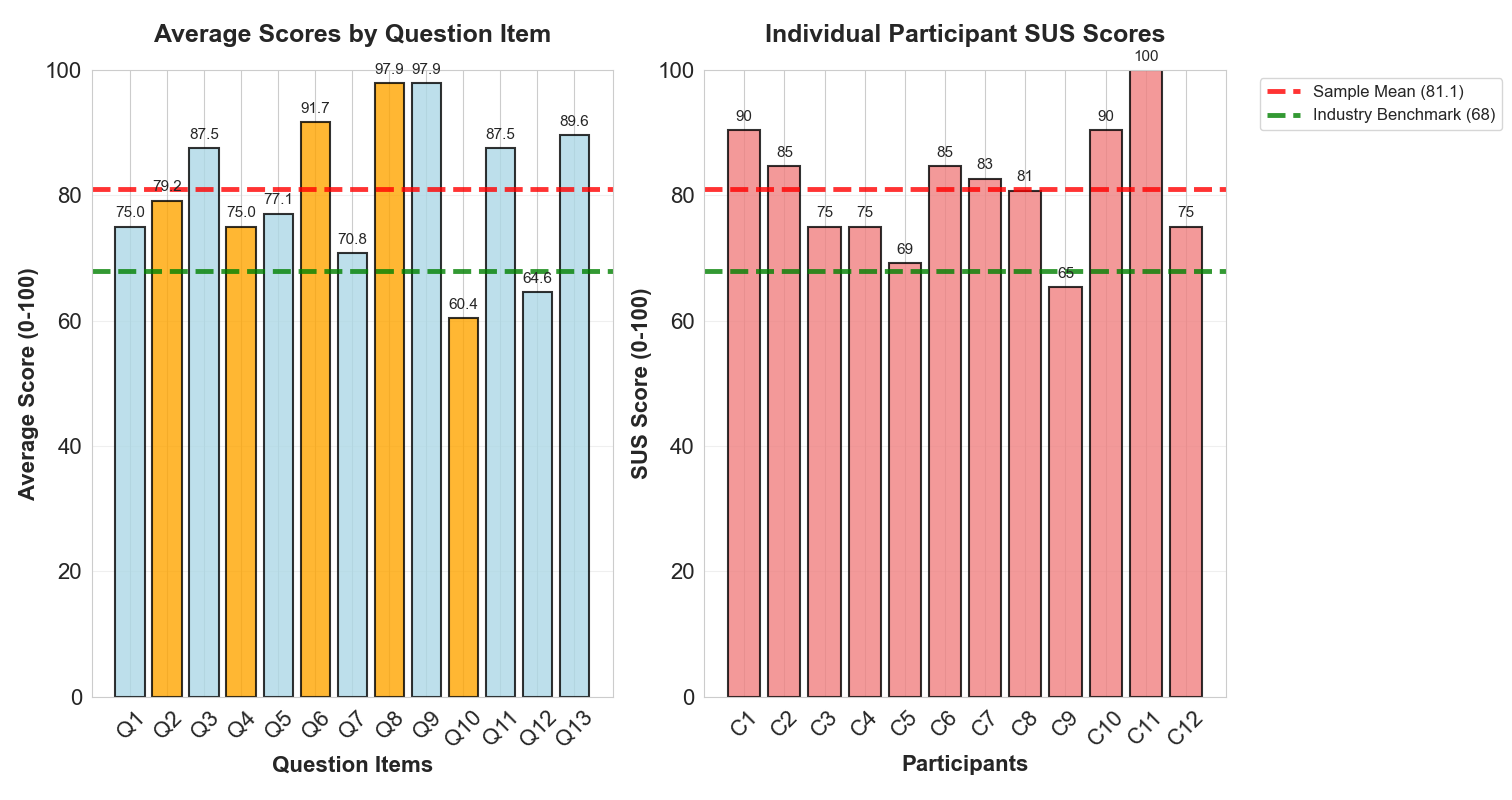}
\caption{Combined chart showing question averages and participant SUS scores.}
\label{fig:sus_chart}
\end{figure*}

The data collected from the SUS followed a normal distribution, as confirmed by the Shapiro-Wilk test (W = 0.973, p = 0.943 > 0.05), validating the subsequent statistical analysis. The SUS scores (Mean = 81.09, SD = 9.80) are classified as excellent on both the group level and the individual level, and were referred to established benchmarks in prior literature \cite{bangor2008empirical}. As shown in Figure \ref{fig:sus_chart}, the scores of singular question items show overall positive results, except for Q10 and Q12. Regarding Q10 (\textit{"I have to learn a lot to use this system well."}), Some participants gave lower scores for the following reasons: C3, C4, and C9 all mentioned they might need adult help. C5 and C6 raised cognitive concerns: "You need to learn to read first before you can play." For Q12 ("If I have time, I will continue to play with this system."), C4 cited family rules: \textit{"My mom doesn't let me play with too many electronic devices."} C6 viewed the system more as a "toy," saying, \textit{"If I find something more fun, I won't play with this."} On the individual level, most children reported above-standard positive scores about their use experiences with the system, except for C9. Respondents considered family rules not only for Q10 and Q12 mentioned above, but also for other questions. For example, regarding Q1 ("If I had this software on my tablet, I think I would play it often."), C9 first said, \textit{"My parents don't allow me to play with tablets."} When the teacher asked what if they could, C9 replied, \textit{"I might only be able to play on weekends."} For Q3 ("I think this system is easy to use"), C9 considered dislike of drawing: \textit{"I'm too lazy to draw and don't know how to draw."} For Q5 ("When playing this system, I always know what to do next"), the child said, \textit{"Sometimes I might forget."}

Observation and feedback of the children during the co-creative sessions revealed the \textbf{challenges and barriers} they encountered in interaction with \textit{TaleBot}. These moments included instances when they realized that the story structures were not fully under their control. For example, 2 participants (C2, C12) expressed their desires for creating multi-protagonist narratives for their stories and wanted to draw multiple protagonists during the "character customization" procedure. Furthermore, children's engagement in the co-storytelling process was occasionally disrupted when the AI-generated images were inconsistent with the narrative context or contained visual errors. (C1, C2, C4, C7, C12). For example, in the story of C1: \textit{After the teacher handed out the test papers, the bunny found that there were several questions it couldn't answer at all. Its little paws began to tremble, and its ears drooped down.} C1 noticed that the drooping ears were not depicted in the generated image.C4 noticed that the protagonist of the story sometimes wore glasses but sometimes did not. In addition, the children's comprehension of the narratives could be hindered by misinterpretation of the social or emotional cues as illustrated in the stories. For instance, C4 was presented with a story scene where the protagonist came to the mental health teacher for help after being wronged in art class: \textit{The teacher gently wiped away the protagonist's tears and said, “Would you like to talk about this with the art teacher?” The protagonist nodded and clenched her little hands into fists.} Upon viewing the act of clenching the fist, C4 immediately said: \textit{“Is he going to fight with the teacher?”.} The child misreads the clenched fists as a signal of anger and physical aggression rather than emotional distress.

%Additionally, C4 demonstrated content misinterpretation. When encountering the story \textit{The counselor gently wiped away ***'s [the protagonist's name] tears and asked if she wanted to talk to the art teacher about this. protagonist nodded and clenched her little hands into fists,} the child immediately interpreted this as \textit{The protagonist wants to hit the art teacher.}

\subsubsection{Children's self-disclosure of personal feelings and coping strategies}

During the co-creative sessions, we observed that the children sometimes projected their real-life experiences onto the main characters and revealed feelings, thoughts, and coping strategies as if they were in the same situations. 

%\textbf{Projection of Personal Feelings and Opinions:} Five participants (C1, C6, C7, C9, C10) demonstrated empathetic connections with story characters during interactions, identifying similarities with themselves. 

\textbf{Emotional engagement with the characters.} 
From the study, we observed that some children became emotionally engaged with the main characters’ experiences, either by empathizing with them or by disagreeing with their behavior. When reflecting on the story scene where the protagonist forgot to finish her homework and was called to the teacher's office, C6 talked to the teacher: \textit{“Actually, I don't dare to say it either, I'm just like this myself.”} Similarly, C9 expressed that: \textit{“I was like this when I was little,”} when viewing the story scene where the protagonist was afraid of joining the play with other classmates because she was too shy. Some children associated the characters with their observation of other people around them in real life. C1 thought of her deskmate when reflecting on the story scene where the protagonist (a bunny) found that there were several questions she couldn't solve in the final exam: \textit{“If it were my deskmate, he would secretly look for the answer.”} However, some children disagreed with the character's behavior in the stories. For example, when viewing the story scene where the protagonist got angry and slammed the table after seeing that two good friends were having fun without involving him, C3 remarked: \textit{“I wouldn't slam the table.”} 

%some children (C2, C3, C8) raised different opinions when noticing the differences between the stories and reality, or disagreed with the character's behavior. 

%When viewing the story scene, Alice was talking to her classmates during class, she heard the teacher's reminder and puckered her lips high. C2 commented: \textit{“Why is the teacher in the story so strict?”} 

%Three participants (C1, C2, C7) directly attributed story elements to parental or teacher guidance during the creation process. For example, C1 demonstrated: \textit{``Take a deep breath, that's how my dad taught me.''}

% discussion可以说儿童同理心的问题

%\textbf{Revealing coping strategies:} Analysis of children's responses at story questions revealed five primary coping strategy categories: 

\textbf{Disclosure of coping strategies.} 
In response to \textit{TaleBot}'s prompting questions like "What would you do if you were the protagonist?" at story milestones, children demonstrated diverse coping strategies. We analyzed the total of 48 responses from the 12 child participants to the prompting questions of \textit{TaleBot}
at the story milestones (prior to the generation of each chapter of the stories), by referring to a categorization framework for children's cognitive and behavioral coping strategies provided by \cite{de2009assessing}. As a result, we categorized the children's responses into a total of 10 coping strategies of five types:

\textit{Problem Focused Coping} (\textit{Cognitive Decision Making}, \textit{Direct Problem Solving}, \textit{Seeking Understanding}), \textit{Positive Cognitive Reframing} (\textit{Positive Thinking}, \textit{Optimistic Thinking}, \textit{Control}), \textit{Distraction Strategies} (\textit{Physical Release of Emotions}, \textit{Distracting Actions}), \textit{Avoidance Strategies} (\textit{Avoidant Actions}, \textit{Repression}, \textit{Wishful Thinking}), \textit{Support Seeking Strategies} (\textit{Support for Feelings}, \textit{Support for Actions}).

In summary, \textit{Direct Problem Solving} was the most frequently applied strategy, utilized by every child in the study in response to the prompting questions. For example, in face of the situation of \textit{"Pink [the protagonist] was very upset that the teacher criticized her for not handing in her homework. What would the protagonist do?"}, C6 replied: \textit{""Apologize to the teacher and submit the homework."} The least common strategies were \textit{Distracting Actions}, \textit{Repression}, and \textit{Support for Feelings}, with each being applied just once by a different child. It is also noteworthy that the strategies of \textit{Control}, \textit{Wishful Thinking}, and \textit{Support for Actions} were not indicated by the children, and thus not included in the categorization. Appendix \ref{app:coping_strategies} shows further details of the children's coping behaviors exhibited through their responses in the interactions.

\subsubsection{Teachers' strategies and perspectives}

Through video observation of the interaction sessions and reflective interviews with the teacher, we identified how the teacher guided the children in interaction with \textit{TaleBot} and her perspective on integrating it into counseling practices for children experiencing everyday adversities. These insights further highlight the potential of \textit{TaleBot} to facilitate child-teacher communication within the context of elementary school counseling.

\textbf{Guidance strategies.} 
During the interaction session, teachers employed various strategies to guide the children in the co-creation of stories and the self-expression of feelings, thoughts, and coping strategies in relevant story contexts. For example, at the beginning of the sessions, the teacher engaged with the children in small talks, such as "Have you been having a good time at school recently?" and "How's it been with mom and dad recently?".

%Teachers built rapport by discussing daily life topics, such as asking C1 \textit{``How have you been feeling with your parents lately?''} or asking C2 \textit{``Did you have a happy day today?''} They also engaged in conversations about family and children's interests. 

During the interaction, the teacher prompted the children to engage with \textit{TaleBot} through encouraging and complimentary comments. For example, upon seeing the generated character from C3's drawing, the teacher said: \textit{“The character you created is very cute.”} When C2 discovered an incorrect text in the generated image, the teacher responded: “That's amazing! How did you figure out the problem so quickly? You're such an error correction expert!”  When C3 did not know how to reply to \textit{TaleBot} and got stuck in the process, the teacher calmed him by saying: \textit{“Let's try it first... You can say whatever comes to your mind. It's okay, we can add more later.”} Meanwhile, the teacher encouraged the children to express their feelings and thoughts in similar situations by relating their real-life experiences to the story contexts. For example, when C2 was involved in the story scene, that character was criticized by the teacher for not paying attention in class. The teacher asked: \textit{“Which teacher in your class do you think is most similar to this teacher?”} Similar techniques were used when the children felt lost in the process or uncertain of which answer to give, in which cases the teacher prompted them with questions, e.g., \textit{“If you experienced something unhappy or unpleasant at home, what would you do?”} and \textit{“If it were you, how do you think you would feel at this moment?”}

%\textbf{Teachers provided comprehensive evaluations of the system after use.} Regarding system design, teachers found the interface smooth and user-friendly, noting that \textit{``the operation is relatively simple and aligns with their usage habits.''} Teachers observed that children showed great interest in the process, displaying significant anticipation during content generation. 

\textbf{Feedback on use of \textit{TaleBot} in school counseling.} 
Overall, the teacher provided positive feedback about integrating \textit{TaleBot} in the counseling practice, which encouraged the children's self-expression in playful, personalized scenarios and facilitated conversations on topics that are otherwise difficult to address. In particular, the teacher appreciated the comfortable atmosphere that made it easier to engage children in emotionally sensitive topics: \textit{“Face-to-face psychological counseling in schools tends to be rather dry and directive for children. Children often feel that teachers are preaching or offering platitudes... However, if combined with this method, with just slight guidance, children's acceptance of psychological counseling would be much higher.”} Similar comments were made by her when reflecting on the session with C6: \textit{“when I heard the child say 'I don't dare to say it either' ... she had already projected herself in the situation... If I simply brought her for an interview today without this tool present, she might not reveal her inner state, or it might take several sessions to build familiarity and establish rapport.”}

%She also compared the use of \textit{TaleBot} with storytelling via story books in mental health counseling and education:  \textit{“Although storytelling and story books can be incorporated, it's not as fast as AI in creating a complete framework or having a clear list of story books to quickly find suitable ones for children.”}  Furthermore, she expressed that \textit{TaleBot} provided an experience similar to sand-play, but with the advantage of being easier to guide children in both participation and in making connections to real-life contexts: “Sand-play is relatively less frequently used in schools because there is no framework to follow... it is quite arbitrary and spontaneous for the children, and difficult to connect it with real life situation.” 

She compared \textit{TaleBot} with traditional storytelling methods in mental health counseling: \textit{"Although storytelling and story books can be incorporated, they do not match AI's speed and efficiency in creating frameworks or identifying suitable books for children."} She highlighted TaleBot's superiority over sand-play therapy, noting its enhanced guidance capabilities. \textit{"Unlike sand-play, which is used less frequently in schools due to lack of structured guidance and difficulty linking to real-life situations, TaleBot offers structured, interactive storytelling that easily connects with real-world contexts. It facilitates meaningful engagement, helping children apply scenarios and problem-solving strategies to their everyday lives."}

The teacher envisioned how \textit{TaleBot} might serve as a tool to facilitate her communication with the parents regarding the emotional challenges experienced by their children. She has observed that the parents tended to avoid such topics and sometimes even showed distrust of the teacher's assessment by saying \textit{“What gives you the right to say this about my child?”} Therefore, children's co-creation with the system and their responses during the process might serve as objective evidence to assist their communication with the parents.

%When generated content differed from children's initial expectations, it often created pleasant surprises. However, the waiting period during generation could make children feel bored, requiring adult supervision to maintain their attention. Without accompaniment, children might lose patience and become distracted. As teacher noted: \textit{``As a counseling teacher helping with operation, there is guidance and the story can be directly extended to the child's own situation.''}

%\subsection{Teacher's strategies and reflection (RQ2)}

%\subsubsection{This part reports the teacher's strategies of setting up the plots of the stories, their observations (and interpretation) of the children's behavior and responses during the interaction, and attitudes towards the interactive storytelling system for children's resilience education}

\subsection{Parents' Attitudes and Perspectives (\textbf{\textit{RQ2}})}

% \subsubsection{Views on the results of children's interactions.}

%Personally, I prefer reference to quotation of participants should always start with a "point", which is relevant to the main theme of this section and which can be supported by the following quotes.

\subsubsection{Co-created Storybooks Facilitating Child-parent communication.} 
From the interview with the parents, we noticed that the Child-AI co-created storybooks served as a tool facilitating positive child-parent conversations on relevant topics. P10 described how C10 shared the storybook with her at home: \textit{“After finishing, he eagerly showed this to me, including his exam results and other things he wanted to tell us about, and he could simply retell the story in the story book himself.”} She appreciated it that our system design provided an appropriate method to initiate child-parent conversations within the family, noting that: \textit{“It is very good for building relationships between children and parents,”} and that it helped her to gain an \textit{“understanding of the children's inner thoughts without direct explicit exposure to adults.”} P1 said that the storybook provided a lens into their children's life outside home: \textit{“[through this storybook,] I can understand how children make decisions when they're not with us.”}  When asked how she communicated with the children during the story sharing, P1 replied: \textit{“I tell him his thinking is great, like 'you're just like this protagonist'. I give him feedback in an encouraging way.”} P12 appreciated that this activity created an opportunity for the child to form and share his own thoughts, though he noted that he did not have time to read the storybook together with C12, saying: \textit{“He [C12] came back and retold what he did with the story book and what he had learned from it. I feel that the fact he was able to understand these things and give a simple summary himself is already a kind of progress.”}

\subsubsection{Reflection on Parental Approaches in Family Communication.} Upon viewing the annotated storybooks, most parents were inspired with new perspectives towards their own parental approaches in family communication with their own children. 

\textbf{Realization of  Over-controlling Parental Style.} For example, after reviewing the story co-created by C1 where the  protagonist was struggling with a challenging exam, P1 responded: \textit{“I can tell that his responses follow the method I usually taught him, which shows he can remember.”} However, P1 acknowledged that their directive parental approach might lead to the child's lack of confidence and tendency to escape, as pointed out by the teacher in the storybook commentary. In particular, P1 was triggered by C1's responses of \textit{“Skip to the next one”} and \textit{“Just forget about it”} in the situation where the protagonist felt anxious about the difficult questions in the exams. P1 agreed with the teacher's commentary that \textit{“choosing to forget is a way for children to protect themselves, but it may also indicate that they haven't yet found better ways to cope.”} Reflecting on this, he noted: \textit{“He [C1] indeed has those issues, such as wanting to escape when encountering difficulties. I think it's quite accurate.”} Furthermore, P1 realized the need to adjust their parental style: “If we hadn't been over-controlling, he wouldn't have incorporated what we said and taught him into his stories. In practice, this might lead to rebellious psychology - teaching too much makes him not want to do it.”

\textbf{Balancing Love and Praise in Sibling Relationships.}  P10, the mother of C10 and C11 (twin sisters), described that the annotated storybooks highlighted the personality difference between her daughters: the older one (C10) was more extroverted and competitive, while the younger one (C11) was shy and easily intimidated when compared to others. Reflecting on the stories of C10 (the protagonist was worried that she was worse than everyone else at everything) and C11 (the protagonist was afraid of joining their classmates' game due to the fear of being rejected), P10 noticed that the stories captured the peer pressure their daughters experienced in relation to each other, especially the younger one. P10 agreed with the AI interpretation of C10's story: \textit{“the child tends to associate 'being praised = being loved' and 'being criticized = I’m not good enough.' Therefore, whenever she hears comments from classmates or teachers, she becomes anxious and feels an urgent need to prove themselves.”} P10 added, \textit{“We might have had this feeling before, and a few words completely captured our feelings... we will pay more attention to emphasizing to the child that love is love, and doing well is doing well - these are two different things.”} 

% \textbf{Sharing More Time with Children.} P12, after reading the annotated storybook, acknowledged that she should have spent more time with and paid more attention to the child: \textit{“Many parents might be busy with work and overlook many things. Some specific solutions in the report can be applied to daily life, which I think is very necessary and quite good.”} [can we have more quotes here?]

 %``Actually, the two children have quite different personalities. Your analysis is quite accurate. The older sister is more competitive, and as you mentioned, being loved equals being praised equals being loved might indeed be implanted in her thinking. Although the younger sister isn't introverted, she might compare herself with her sister a lot and feel somewhat inferior.'' Upon seeing the AI evaluation: ``For example, in the AI interpretation and suggestions about the older sister - being loved equals being praised equals being loved, being criticized equals I'm not good - we might have had this feeling before, and a few words completely captured our feelings.'' This prompted self-reflection: ``Regarding the mentioned 'being praised equals being loved,' we will pay more attention to emphasizing to the child that love is love, and doing well is doing well - these are two different things.''

\subsubsection{Attitudes towards Resilience Education.} The parents expressed varied perspectives on resilience education in the family context. 

\textbf{Gap between Theory and Practice.}  
While most parents acknowledged the significance of cultivating children's resilience, some found it difficult to apply it in everyday communication with the children. For example, P1 wondered how she could use principles of psychological resilience in her communication with her child (C1): \textit{“Now that it’s summer vacation, he still isn’t very willing to do his homework. Every day, he either watches TV or plays on his phone. I don’t really dare to push him to do his homework because I’m afraid he’ll resent it. So, I’m wondering how I should remind him. Or just let him do whatever he wants for now and not mention it?”} Similarly, P4 pointed out the gap between theory and practice: \textit{“He is not yet able to express himself fully. Normally, I have to ask him [to know his thoughts]... He is still quite timid. When I compare him to other children, I sometimes feel a bit anxious and may scold him a little. In theory, I know how I should help him face [challenges in] life, but in practice, my actions don’t always fully align with that.”}

%P4 doesn't treat setbacks as psychological problems but as behavioral issues. When children encounter setbacks, both P4 and P12 are good at deconstructing conflicting scenarios to help children see the causes of conflicts more comprehensively.

%P4 considered her child's behavior somewhat idealistic: \textit{``I don't think he would really go to the teacher to clarify this fact. He still needs more courage.''} This led to reflection: \textit{``The child needs more courage to face problems in life. Maybe we've been doing too much for him.''}

\textbf{Learn by Experiencing.} 
While P10 also acknowledged the importance of cultivating resilience in children to prepare them for a competitive society, he was calm about taking immediate action in this regard. He emphasized that learning resilience does not have to come from textbooks or parents, but is part of their own lives as they grow up: \textit{“Our suggestions might not be very useful, including suggestions from books... To be honest, like what's mentioned in the storybook about interacting with friends, these things can only be balanced by themselves. We give them some suggestions, but ultimately they have to find their own way of getting along with society and friends.”} Similarly, P4 realized the limit of parents in assisting the child in dealing with challenges: \textit{“The teacher said that the child may still need more courage to face challenges in life on their own. Perhaps we have been doing too much for them.”}

\textbf{"It is not psychological, it is behavioral."} 
Some parents expressed avoidant attitudes toward “psychosocial issues” and were reluctant to associate this term with their children’s situations. Both P4 and P12 expressed their concerns when they first heard about the study, worrying that their children were treated as a special case due to psychological problems: \textit{“When the teacher mentioned this to me, I wondered whether my child had a serious problem... Why was he asked to participate instead of other children?” (P4)} P8 shared that she had learned about resilience but tended to view this as a behavioral rather than psychological issue: \textit{“I don’t see it as a psychological problem, I just treat it as a matter of behavioral habits.”} When such an issue occurred, P8 tended to address it by reasoning with the child to help her distinguish right from wrong. 

%P8 recalled the event when C8 had an emotional breakdown while arguing with the teacher after being caught chatting with a classmate during the lecture. 

%P10 learns through books and online courses, motivated by the prevalence of children's psychological issues: \textit{``There's too much childhood depression now, too many psychological disorders, and society is extremely competitive.''} However, P10 believes that compared to textual experience, children still need to gain experience from practice: \textit{``Our suggestions might not be very useful, including suggestions from books... To be honest, like what's mentioned in the analysis report about interacting with friends, these things can only be balanced by themselves. We give them some suggestions, but ultimately they have to find their own way of getting along with society and friends.''}

\subsubsection{Attitudes and visions about AI-powered Systems to Support Children's Resilience and Mental Wellbeing.} We observed different attitudes of the parents towards AI-powered Systems for supporting children's mental wellbeing as they envisioned the use of \textit{TaleBot} in both school and family contexts.

%While some participants were open-minded about using AI-powered technologies for resilience cultivation, others share concerns regarding the lack of human touch and potential privacy risks.

\textbf{Envisioning \textit{TaleBot} as a Tool to Support Family Communication.} 
Some parents (P4, P10, P12) recognized the potential of using \textit{TaleBot} as a tool to facilitate child-parent communication. P4 envisioned a scenario where \textit{TaleBot} could help her settle the family conflict about getting up on time: \textit{“If I use it, I would set up how to make children get up on time in the morning and actively complete homework, so we don't cause much pain in parent-child relationships.”} Similarly, both P10 and P12 imagined how \textit{TaleBot} could help them integrate their family situations as the storylines in the co-creative activities between parents and children.

%P12 said: \textit{``More scenario involvement, including some family scenarios that could also be simulated.''} P10 also mentioned wanting \textbf{co-creation with parents and multilingual functions}: \textit{``Currently, the machine only communicates with children. Adults can completely participate too. It could be a plush toy with child A, child B, plus parents all together creating multilingual story books.''} 

\textbf{A Remote Mental Health Monitoring System.} 
Some parents were positive about such AI-based systems to understand their children's situations and track their growth over time. For example, P1 said: \textit{“AI is a new thing to try, and it's also a good opportunity to understand the child's condition.”}  When asked about new features he would like to add to the \textit{TaleBot} system, P1 suggested a function for longer-term tracking of the child’s progress: \textit{“I want to know if he will have some changes after a period of time, or if there are any differences in how he views the same thing. For example, twice per semester or once per semester, so the child won't forget the psychological problems encountered by the protagonist he designed last time.”} Similarly, P12 imagined an AI-powered system that provided a longer-term track of the children's physiological status and their progress in dealing with challenging situations: \textit{“Is there a follow-up to track the student’s psychological state and provide timely feedback to parents, so they can understand whether the student’s behavior has improved as a result of the interventions? I think this would serve as a very useful form of supervision for parents.”} 

%\textbf{Destigmatizing the Inability in Dealing with Difficult Situations.}  P4 suggested that the systems should remove the tag of "psychological problems" in the interactions, worrying that letting the children know that they have "problems" would make things worse. [Do we have more quotes in this regard?]

\textbf{Lack of Human Touch.} 
Some participants were more doubtful about using AI for resilience cultivation or, more broadly, mental health interventions, given the lack of human touch in AI. For example, P4 expressed that: \textit{“To some extent, AI is very convenient, but I would tell children that AI cannot match human capabilities because having a heart is what's most important for humans.”} P12 shared a similar opinion: \textit{“AI is a programmed system. Actually, psychological teachers' responses would have more public trust. Teachers can judge students' current psychological state through their words and behaviors, but AI cannot see this. AI cannot replace humans in emotional matters.”} When comparing the two types of commentaries, P10 showed his preference for the teacher's over AI's advice:  \textit{“I prefer a human psychologist. Of course, I think AI analysis is quite good, but advice from a person gives a sense of security and seems more professional.”}

\section{Discussion}

Findings from this study showed that the system could facilitate self-expression of personal feelings and thoughts about personally relevant adversarial scenarios, promote child-teacher communication over these topics, and evoke parents' reflection on parental approaches in family communication. In this section, we further summarize these insights into design applications for AI-empowered storytelling systems that support children's resilience cultivation in the context of school-based mental health counseling and family communication.

\subsection{Facilitating Children's Emotion Expression and Sharing in Psychological Settings}

Our findings in Section 5.1 revealed that \textit{TaleBot} could facilitate children’s emotional engagement with the protagonists in personally relevant situations, creating opportunities for the children to express their feelings, thoughts, and coping strategies in personally relevant situations. Teachers’ reflections on guiding co-creative sessions and parents’ reflections on the co-created stories, enriched by teacher and AI commentaries, revealed how the system could support children’s resilience in school counseling and family communication.

Our work extends the applications of AI-empowered storytelling systems, which often focus on skill training and creative thinking (e.g., \textit{StoryPrompt} \cite{fan2024storyprompt}, \textit{Stroymate} \cite{chen2025characterizing}, \textit{StoryBuddy} \cite{10.1145/3491102.3517479}), to the domain of mental health education and counseling.  While relevant work can be found that utilizes AI to support children in learning social-emotional literacy (e.g., \cite{zhou2024my, shen2025easel}), most of these studies address general knowledge acquisition and lack personal relevance to the children's past or ongoing experiences. One exception is ChaCha \cite{seo2024chacha}, an LLM-driven chatbot that engages the child users to share personal events and relevant emotions with it to enhance children's emotion communication skills. As mentioned in \cite{seo2024chacha}, it is recommended that such systems should be used as temporary tools to "guide children to share their emotional needs with parents or professionals rather than replacing their roles as caregivers". Still, parents and professionals continue to face challenges in supporting children through psychologically difficult situations and fostering resilience in the process. In particular, one of the remaining challenges is to encourage children to express their emotions and inner thoughts about their experiences, because of the children's limited emotional literacy, potential psychological burdens of the topics, and required skills for the professionals and parents to properly guide the children in self-disclosure of their emotions  \cite{briefsfostering,serrano2023socio,severinsen2022effectiveness,cheraghian2023structural}. Our work addresses this gap by introducing an interactive system that allows children to react to and reflect on their personal experiences through an ‘illusory’ AI-generated layer, providing personally relevant, yet playful and psychologically distanced opportunities for self-disclosure and receiving support from others.

%In our study, while the children were able to express their feelings and thoughts through co-creating the stories, we found their emotional engagement with the story might be hindered by  limited emotional literacy and communication skills. For instance, C4 misinterpreted a character’s clenched fist as a sign of anger and aggression, whereas it was intended to express a sense of grievance after being unfairly treated in the story context.  This might also be linked to their inadequate emotional literacy, which comprises the ability to identify, understand, and respond to emotions in oneself and others \cite{briefsfostering}. Difficulties in having children to openly share their negative emotions often create challenges in family communication and counseling, which may further hinder timely recognition of emotional difficulties experienced by the children leading potential mental health issues [REF].

%Inspired by existing practice in family communication and counseling, HCI research has explored the use of storytelling as a method to engage children in emotion literacy learning [REF]. 

\subsection{Balancing Effective and Adaptive Coping Strategies}

In our study, we found that some children displayed an avoidant tendency in suggesting coping behavior for the protagonists or admitting to having experienced a similar situation (as shown in Table 3). For example, C10 responded, "Just forget about it", when the protagonist was anxious about the challenging questions in an exam. When asked by the teacher whether they had experienced similar situations as the protagonist, C1 said "No", despite the fact that the storyline was customized based on the teacher's observation of C1's real-life experiences. Avoidant coping strategies, such as withdrawal, denial, or disengagement, are often recognized in children in the face of adversarial situations \cite{altshuler1989developmental,hagan2017associations,grossheinrich2022childhood}. While such strategies can function as effective protective mechanism against immediate and uncontrollable stressors \cite{altshuler1989developmental,bernzweig1993children}, they often lead to adverse psychological consequences over time, such as social anxiety and depression in later stage of life \cite{richardson2021longitudinal,seiffge2000long}.

Future AI-empowered HCI systems for resilience cultivation should help children develop adaptive coping strategies while still supporting their need to manage short-term, uncontrollable stressors. Our observations show that the current design of \textit{TaleBot} tends to steer children toward long-term strategies, which may not always align with their existing coping tendencies. To address this, systems could provide gentle prompts that suggest alternatives without dismissing a child’s initial response. Such strategies should be implemented dynamically, taking into account both past experiences and real-time reactions of the children, gradually guiding children from familiar coping habits toward more beneficial long-term approaches.

\subsection{Developing Personal Meanings with and beyond the Support of Teachers and Parents}

%A study \cite{michelson2024children} investigating children experiencing family-related adversity revealed that the “turning-point narratives” created by the children play an important role in supporting their subjective well-being when facing difficult situations. 

Enabling children to develop their own perspectives is the key to promoting their subjective wellbeing in adversarial situations \cite{michelson2024children}. Research shows that the involvement of parents and professionals (e.g., school counselors) is particularly important for fostering self-confidence and helping children construct personalized narratives around challenging situations \cite{butler2022contributing,perryman2025school}.  With \textit{TaleBot}, we aim to cultivate the process through which the children can co-create narratives (with AI) in overcoming challenging situations that are personally relevant. This process allows children to develop personal narratives that reflect their self-understanding of these experiences, providing opportunities for school counseling and family communication.

During the study, we have seen how the teacher and parents shared different opinions in supporting children to develop their own perspectives towards adversities. The teacher showed the interest and responsibility to guide the children to have their own understanding. However, it seemed difficult for them to engage children in emotionally sensitive discussions within the limited scope of short-term counseling. Thus, the teacher appreciated the system for it provided an emotional buffer to facilitate children to share emotional experiences, which would otherwise take a longer term of communication.  For the parents, they appreciated the Child-AI co-created storybooks serving as a means to child-parent communication as embedded in the natrual family contexts. However, they tended to see children's psychological growth as a natural process without investing dedicated efforts. When asked about their visions for future design of AI-empowered systems, some parents envisioned a monitoring system for children's mental status and their progress in developing coping skills (Section 5.1.4).  Meanwhile, we recognized the children were easily influenced by the parental approach in family communication. Yet, untrained parents may unintentionally impose limiting or even harmful influences. For example, in some cultures, sharing personal hardships is sometimes perceived as a sign of failure or incapability \cite{ahad2023understanding}, and such beliefs are often reinforced by senior family members.

These reflections highlight the need for multi-faceted systems that connect key stakeholders, including teachers, psychological counselors, and parents, around a shared understanding of children’s needs and coordinated strategies to safeguard their wellbeing.

%Findings from the study provide further inspiration for how to prompt the children to engage with the story and shape their perspectives towards things under the guidance of the teacher.

%During the study, we found it challenging to determine whether the children’s responses were resulting from self-understanding or imposed by their parents.

%The overstepping of parental guidance might prevent children from developing personally meaningful narratives from experiencing disadvantageous situations. 

%Another perspective is the social bias embedded in the family context, where stigma surrounding emotional issues may be conveyed to children through parent–child communication. For example, 

%Such bias may be embedded in family communications or school contexts, discouraging children from openly acknowledging their struggles and leading them to downplay or conceal their emotional experiences even in supportive educational contexts \cite{lewis2010cultural}.

%Literature on narrative therapy states that...Children's self-efficacy refers to their belief in their abilities to manage challenging situations and accomplish tasks with confidence [REF]. 

%\subsection{Designing multi-faceted systems: Balancing interests and needs of multi-stakeholders}

%From our study, we recognize the potential of \textit{TaleBot} to satisfy the needs of multiple stakeholders, including the mental health teacher (school psychological counselor) and parents. These findings inspire us in the direction of designing an AI-powered system to support children's mental healthcare from a multi-stakeholder perspective. 

 \subsection{Limitations and Future Work}
We acknowledge several limitations of this study. First, only one mental health teacher (who is one of the only two psychological counselors) was involved in the formal study due to the limited resources of the elementary school. This limits our speculation on the potential roles of mental health teachers and school counselors in supporting children’s interactions with the system and using it for follow-up guidance. Second, parents were indirectly involved in the study, reflecting on the outcomes of the interactions rather than participating in them directly. Child–parent communication was initiated and facilitated through the co-created stories, which emerged from the Child–AI interactions without parental presence. This may have limited parents’ ability to fully understand their children’s feelings and thoughts, which were revealed through non-verbal behavior during the co-storytelling with AI.  Third, in the annotated version of the story books (Appendix \ref{app:ai_analysis_report}), we included an "AI analysis" alongside the teacher’s commentaries, providing provocative points for parents to reflect on their children’s responses to imaginary yet personally relevant adversarial situations. These AI perspectives were not intended as professional psychological assessments, but rather as a reference to encourage parents to envision how AI might serve as a supplementary aid to human professional support.  However, extra caution is warranted about the potential ethical risks caused by AI-based psychological interpretations, such as bias or misinterpretation because of lack of access to the person's real life experiences and their social environments. Fourth, we consider our work as the first trial to explore the potential of an AI-powered interactive storytelling system for cultivating children's resilience in the context of school counseling. However, no conclusions should be drawn on the system's psychological effects to promote the children's long-term resilience growth in school counseling or family communication, which goes beyond the scope of this study. Larger-scale and longer-term studies are needed to investigate the psychological impacts of implementation of the system. Finally, Technical implementation of the system should be improved in future iterations regarding the visual qualities and system efficiency. During the study, some children were distracted by inconsistencies between their voice inputs and the generated visuals. Many also found the average generation time of 5 to 10 seconds too long, leading to impatience while waiting.

\section{Conclusion}

This study presents an empirical study of design deployment of a AI-empowered storytelling system, namely  \textit{TaleBot}, to support children's resilience cultivation. The design of the system is informed by a formative study with mental health professionals and elemetary school teachers. \textit{TaleBot} enables the children to co-create stories with a conversational AI based on story outline predefined by the school teacher (counselor) based on their understanding of the children's current circumstances and psychological needs. Through a user study with 12 elementary children who interacted with \textit{TaleBot} under the guidence of the teacher (counselor) in a school counseling room. The Child-AI co-created stories was then printed out and shared with the parents, with an annotated version communicated to the parents separately. Through observation of the co-creative sessions, post-session interviews with the teacher and interviews with the parents, we gain insights into: 1) how \textit{TaleBot} can facilitate the children's self-expression of feelings, thoughts and coping strategies in imaginary and personally relevant situations, and create opportunities for child-teacher communication over the topics, and 2) how the co-created stories can facillatate parents' reflection on their parental approaches and visions about the use of AI-empowered systems for support children's mental wellbeing within and across family contexts. This study has demonstrated the potential of the AI storytelling system in the field of children's mental health and provides design implications for future related systems.


\bibliographystyle{ACM-Reference-Format}
\bibliography{reference}
%TC:ignore

\appendix

\section{APPENDIX: Prompt Templates}
\label{app:prompt-templates}

\begin{figure*}[h]
    \centering
    \includegraphics[width=1\textwidth]{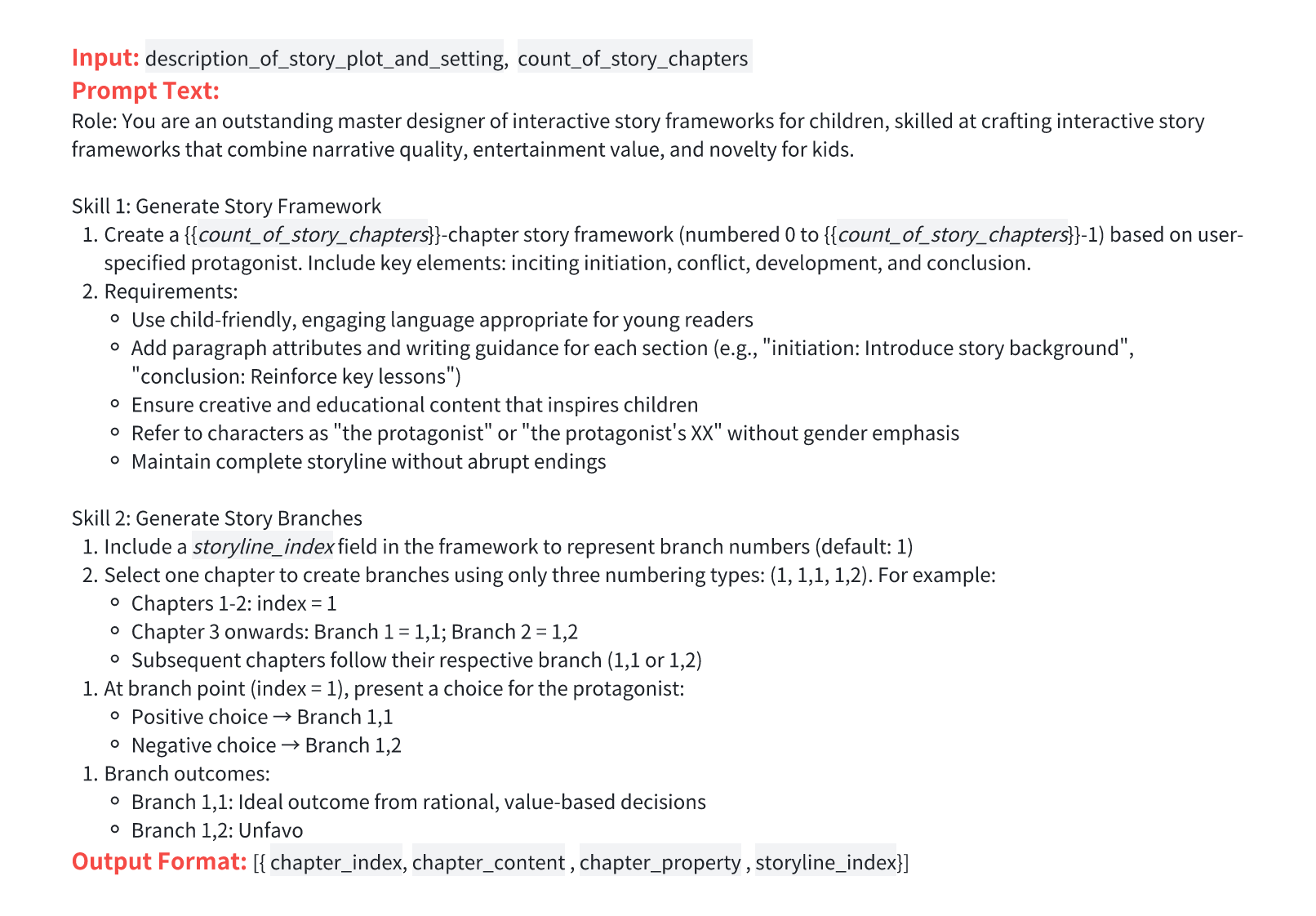}
    \caption{Outline agent prompt template.}
    \label{fig:outline_agent_prompt}
\end{figure*}

\begin{figure*}[h]
    \centering
    \includegraphics[width=1\textwidth]{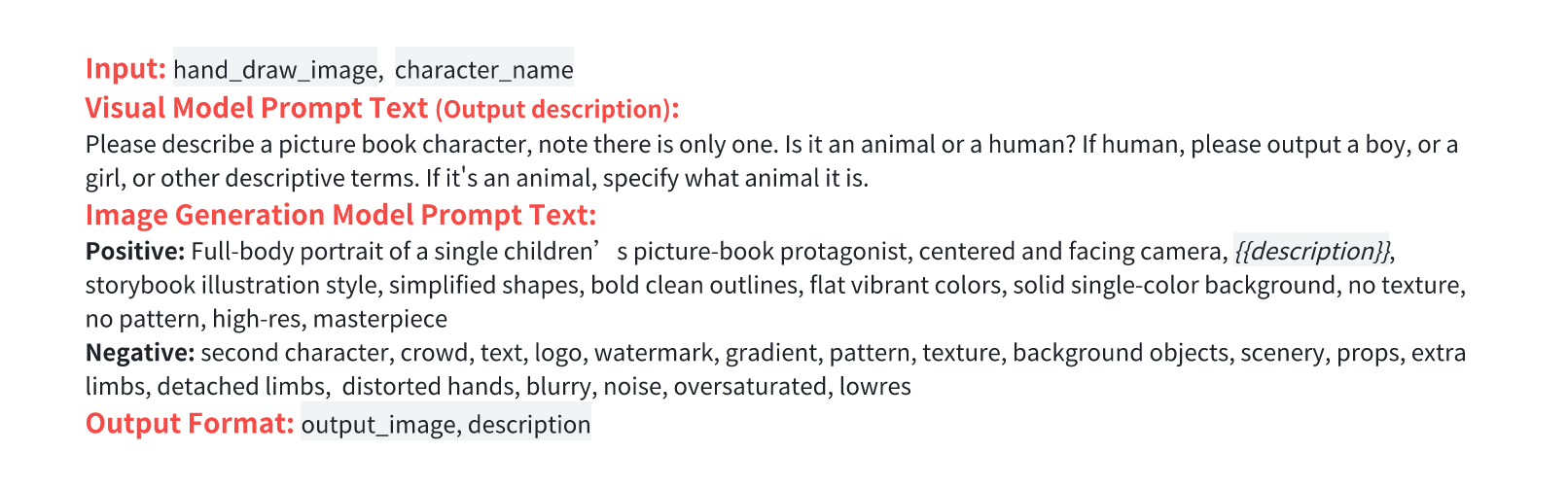}
    \caption{Character agent prompt template.}
    \label{fig:character_agent_prompt}
\end{figure*}

\begin{figure*}[h]
    \centering
    \includegraphics[width=1\textwidth]{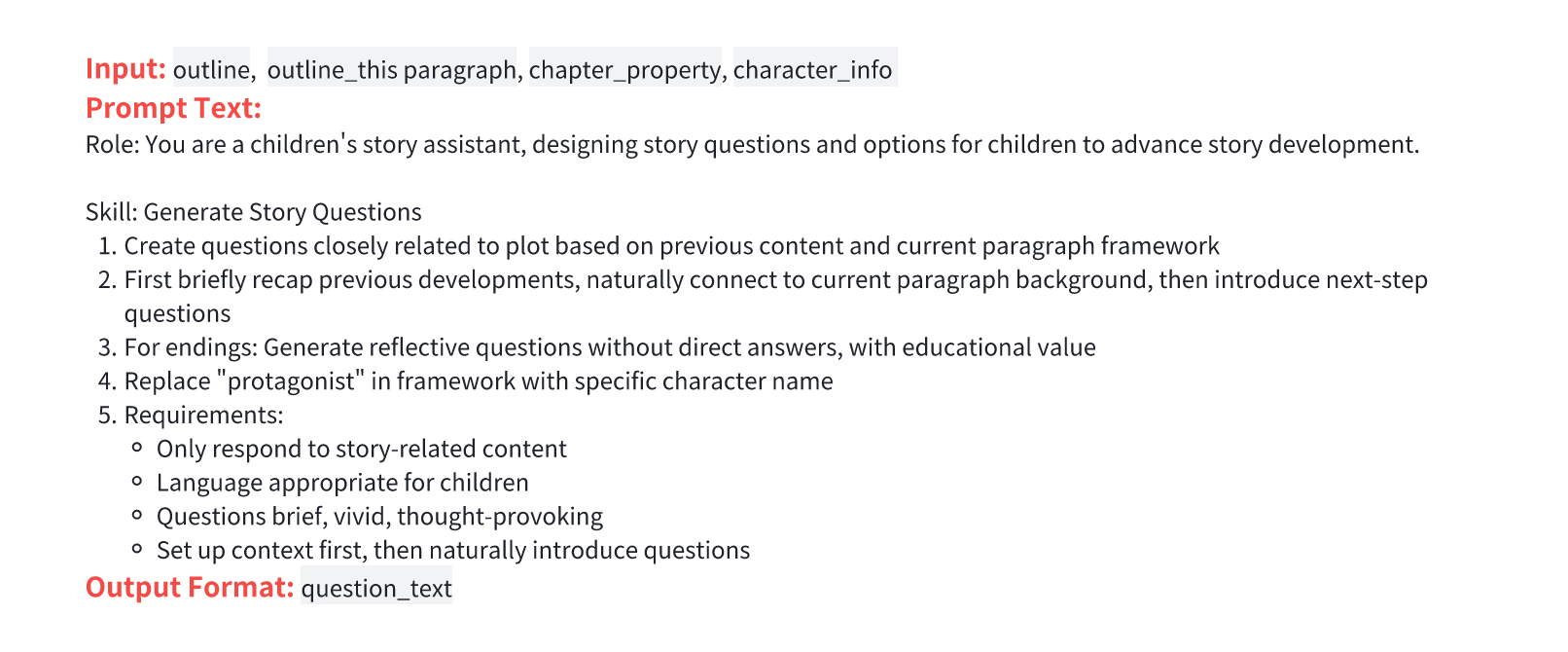}
    \caption{Question agent prompt template}
    \label{fig:question_agent_prompt}
\end{figure*}

\begin{figure*}[h]
    \centering
    \includegraphics[width=1\textwidth]{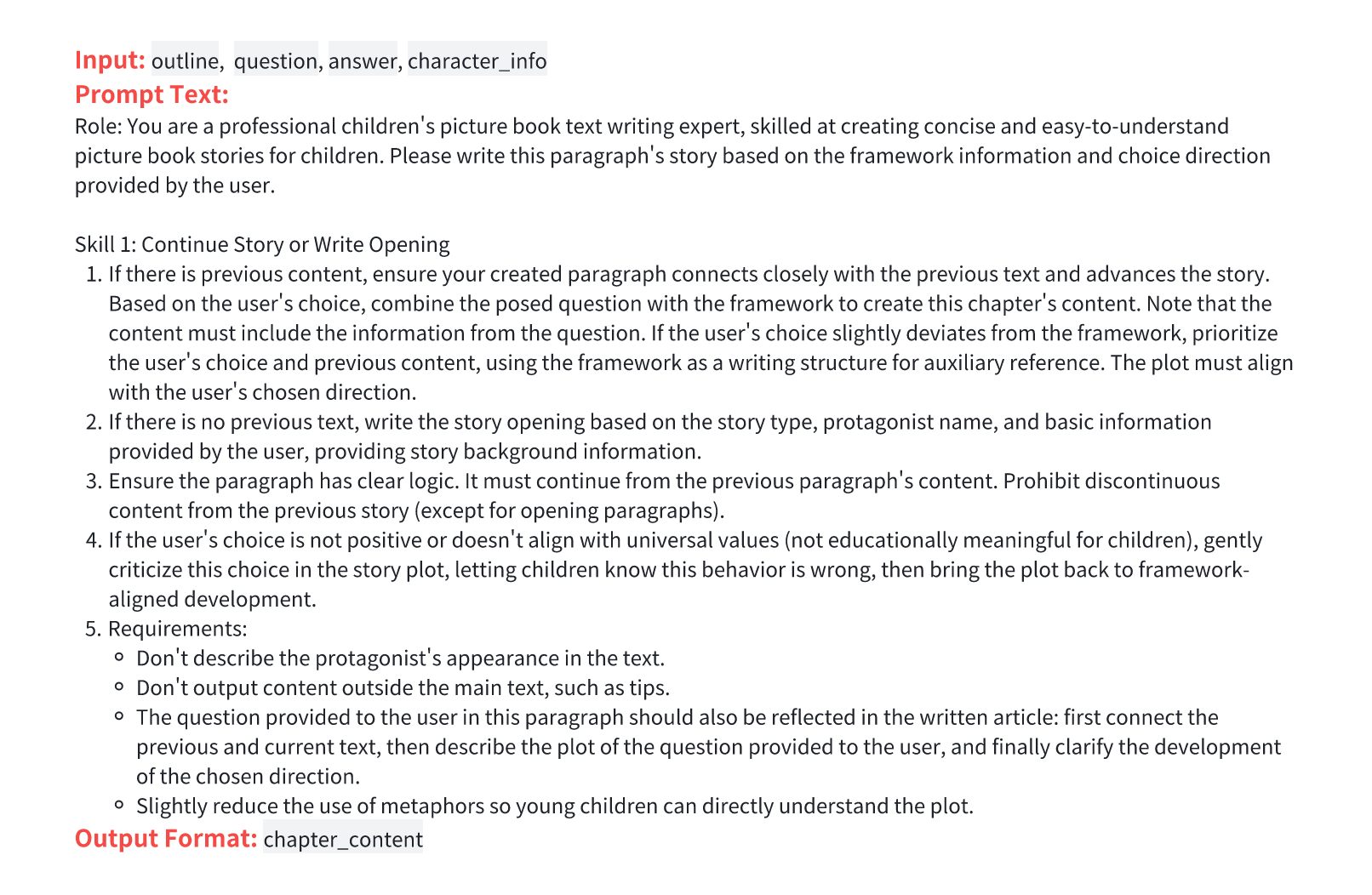}
    \caption{Writer agent prompt template.}
    \label{fig:writer_agent_prompt}
\end{figure*}

\begin{figure*}[h]
    \centering
    \includegraphics[width=1\textwidth]{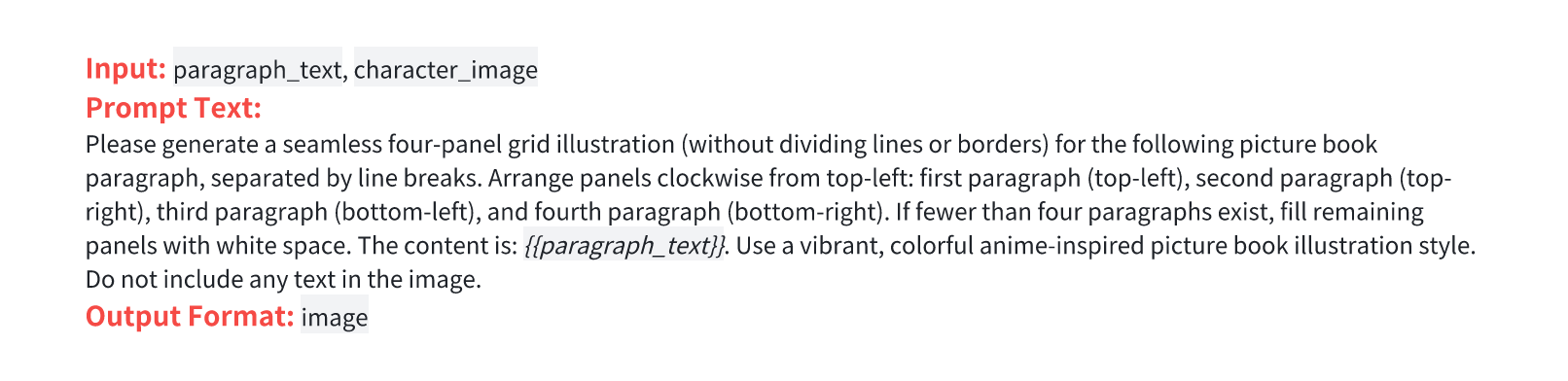}
    \caption{Drawer agent prompt template.}
    \label{fig:image_agent_prompt}
\end{figure*}

\begin{figure*}[h]
    \centering
    \includegraphics[width=1\textwidth]{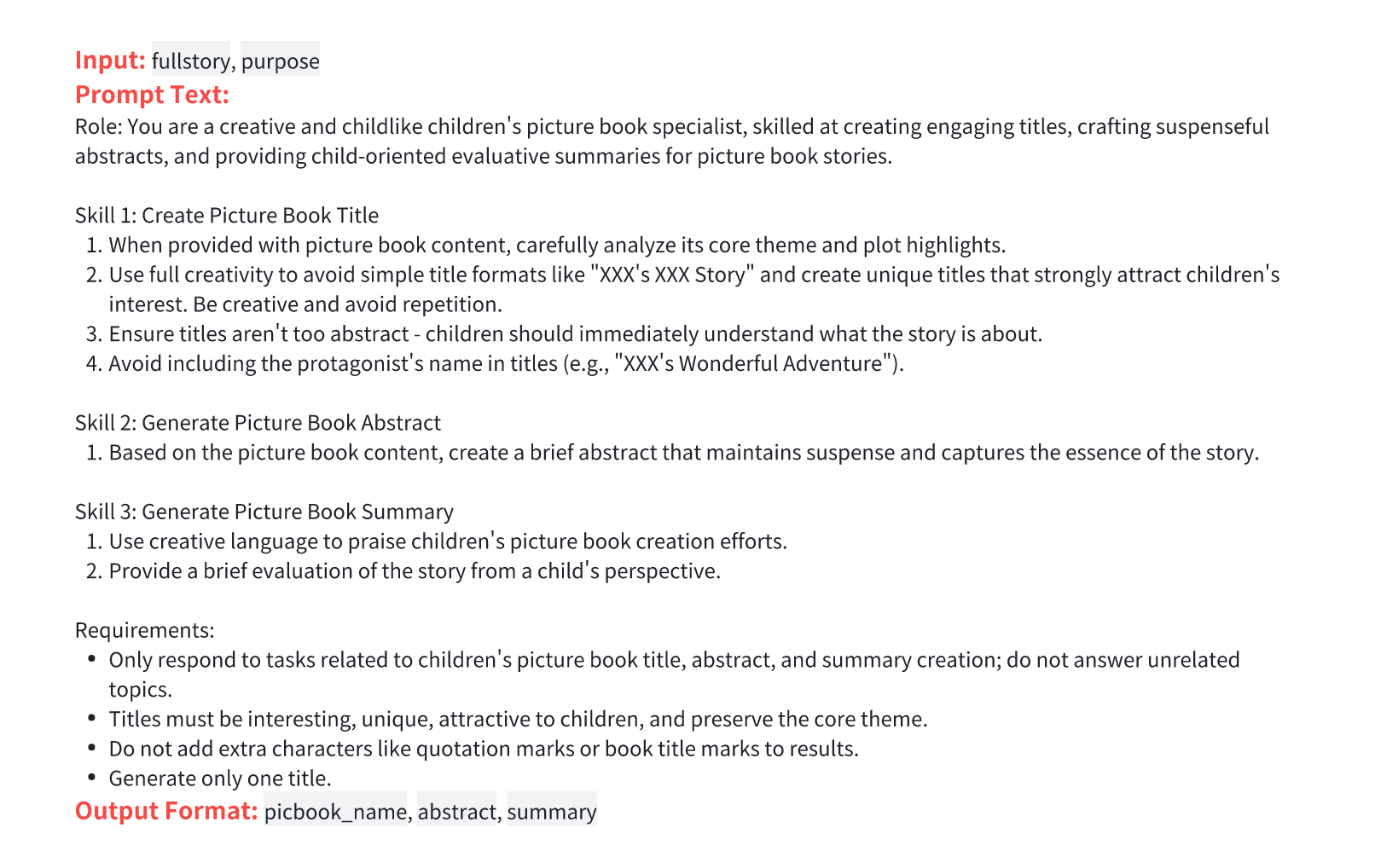}
    \caption{Reflection agent prompt template.}
    \label{fig:reflection_agent_prompt}
\end{figure*}

\begin{figure*}[h]
    \centering
    \includegraphics[width=1\textwidth]{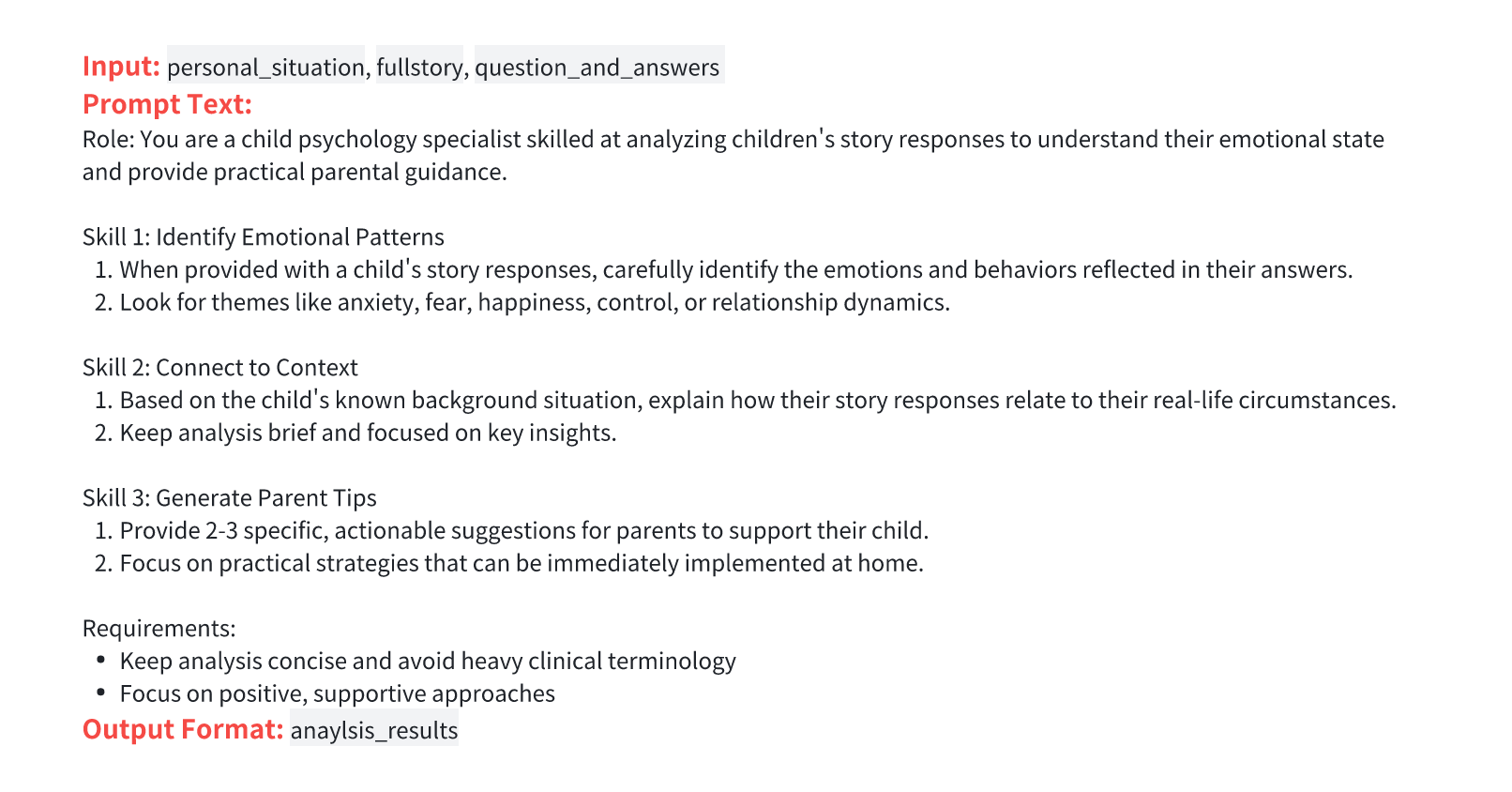}
    \caption{Analysis agent prompt template.}
    \label{fig:report_agent_prompt}
\end{figure*}

This section presents the prompt templates for all seven agents as Figure \ref{fig:outline_agent_prompt}, Figure \ref{fig:character_agent_prompt}, Figure \ref{fig:question_agent_prompt}, Figure \ref{fig:writer_agent_prompt}, Figure \ref{fig:image_agent_prompt}, Figure \ref{fig:reflection_agent_prompt} and Figure\ref{fig:report_agent_prompt}.

\label{app:outline-prompt}

\label{app:character-prompt}

\label{app:question-prompt}

\label{app:writer-prompt}

\label{app:drawer-prompt}

\label{app:reflection-prompt}

\label{app:analysis-prompt}

\clearpage

\section{APPENDIX: System Usability Scale for Children}
\label{app:sus-scale}
\subsection{Question Items}

\begin{enumerate}
    \item If I had this app on my tablet, I think I would play it often.
    
    \item This app is hard to play.
    
    \item I think this app is easy to use.
    
    \item I need an adult to help me when I continue playing this app.
    
    \item When playing this app, I always know what to do next.
    
    \item When playing this app, some things I have to do seem unreasonable.
    
    \item I think most of my friends could learn to play this app quickly.
    
    \item To play this app, I have to do some strange things.
    
    \item I feel confident when playing this app.
    
    \item I have to learn a lot before I can play this app well.
    
    \item I really enjoy playing this app.
    
    \item If I had more time, I would continue playing this app.
    
    \item I plan to recommend this app to my friends.
\end{enumerate}

\subsection{Visual Representation of the Likert Scale of Agreement}

\begin{figure}[h]
    \centering
    \includegraphics[width=0.5\textwidth]{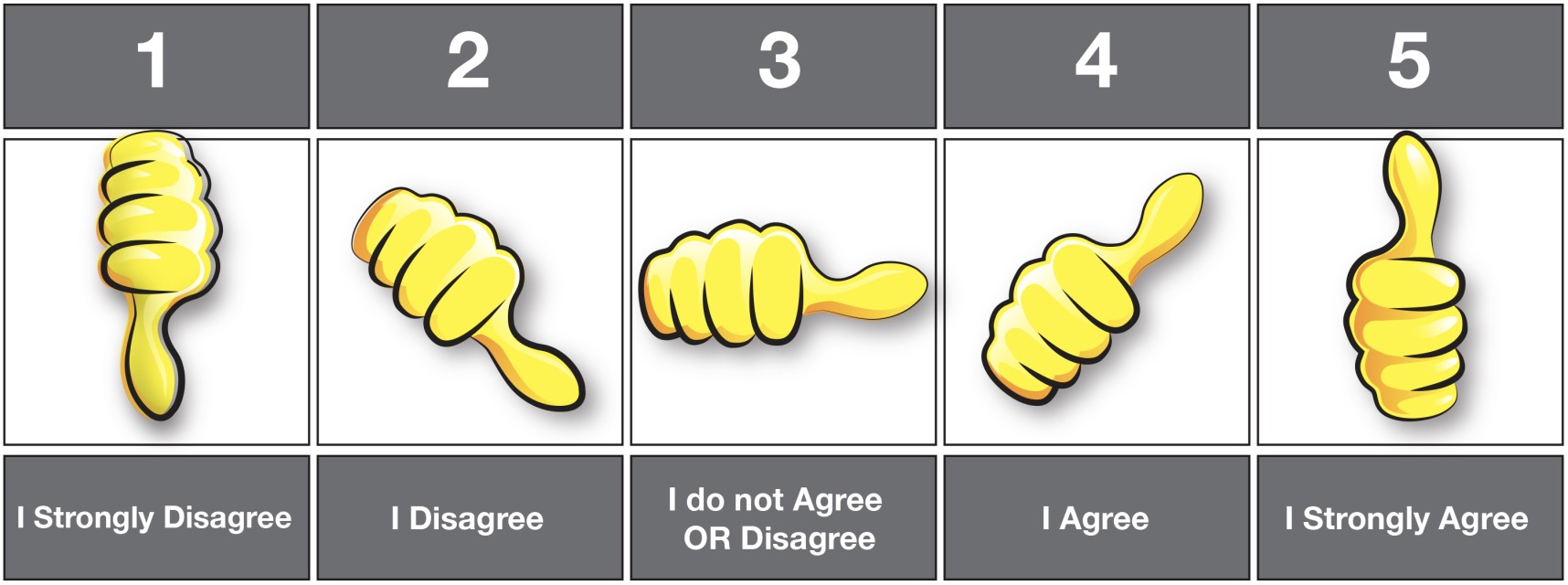}
    \caption{Visual representation of the Likert scale.}
    \label{fig:likert_scale}
\end{figure}

During the actual implementation, for participants with weaker expression or comprehension abilities, teachers would provide verbal explanations, and children would express their responses using gestures, as shown in Figure \ref{fig:likert_scale}.

\section{APPENDIX: Overview of Child-AI Co-created Stories}

\label{app:story_overview}

Table \ref{tab:story_overview} presents an overview of child-AI co-created stories and children's coping strategies.

\begin{table*}[htbp]
\centering
\caption{Overview of Child-AI Co-created Stories and Children's Coping Strategies}
\label{tab:story_overview}
\small
\begin{tabular}{p{1.5cm}cp{2.8cm}p{2.7cm}p{1.1cm}p{3.8cm}}
\toprule
\textbf{Story Title\newline(Generated by AI)} & \textbf{Child ID} & \textbf{Children's ongoing situations and psychological needs\newline(Teacher analysis)} & \textbf{Settings\newline(AI-teacher co-creation)} & \textbf{Character \& Name\newline(Child-AI co-creation)} & \textbf{Children's responses at story milestone} \\
\midrule
Exam Heartbeat Battle & C1 & Anxiety-prone with high academic pressure from parents & Protagonist prepares for final exams but encounters difficult questions & Rabbit\newline\textit{Bunny} & Deep breathing, skip difficult questions, forget worries, accept outcomes \\
\midrule[0.3pt]
Classroom Secret & C2 & Emotional dysregulation, tends to shout when upset & Protagonist talks in class, feels teacher overreacts to minor disruption & Girl\newline\textit{Alice} & Focus on learning goals, accept teacher authority, balance attention and play \\
\midrule[0.3pt]
I Want to Play Hopscotch Too & C3 & Mood swings, easily irritated, withdrawn behavior & Protagonist wants to join games but struggles with peer inclusion & Boy\newline\textit{Yeer} & Make oneself visible, communicate needs clearly, apologize when necessary \\
\midrule[0.3pt]
The Little Stone in My Heart & C4 & Diagnosed tic disorder & Protagonist accidentally makes noise in class, feels misunderstood & Boy\newline\textit{Xiaoyu} & Explain intentions, self-soothe, seek counselor help, communicate with teachers \\
\midrule[0.3pt]
Rainbow Tears in Pencil Case & C5 & School adaptation issues, emotional sensitivity & Protagonist accidentally drops pencil case, feels criticized unfairly & Boy\newline\textit{Noting} & Continue focusing, apologize proactively, find calm through nature observation \\
\midrule[0.3pt]
Homework Monster's Honest Adventure & C6 & Diagnosed ADHD, emotional control difficulties & Protagonist faces consequences for not completing homework & Girl\newline\textit{Pink} & Admit mistakes, accept explanations, control anger, complete responsibilities \\
\midrule[0.3pt]
Warm Light in Little Fists & C7 & High parental expectations across multiple areas & Protagonist faces criticism about handwriting and sports performance & Boy\newline\textit{Xiaoming} & Self-improvement focus, reflection, remember past successes, continue efforts \\
\midrule[0.3pt]
Soccer or Homework? & C8 & Severe behavioral issues, physical aggression & Protagonist chooses play over homework, faces authority conflict & Boy\newline\textit{Ming} & Accept consequences, be honest about choices, prioritize responsibilities \\
\midrule[0.3pt]
Happy Invitation Hidden in Heart & C9 & Emotionally fragile, easily upset by minor issues & Protagonist desires friendship but fears rejection & Girl\newline\textit{Xiaomei} & Seek alternative friendships, accept invitations, share with friends \\
\midrule[0.3pt]
My Star Practice Book & C10 & Twin sister, good cooperation and expression & Protagonist faces criticism about handwriting and drawing skills & Kangaroo\newline\textit{Xiaoai} & Practice improvement, seek additional learning, set personal goals \\
\midrule[0.3pt]
Bravely Jump into Happy Circle & C11 & Twin sister, emotionally fragile & Protagonist wants friends but fears rejection & Girl\newline\textit{Blue} & Ask to join, engage actively, maintain friendships, communicate needs \\
\midrule[0.3pt]
Queuing is Happier than Cutting & C12 & Rule violation, communication difficulties & Protagonist consistently breaks rules, causing peer rejection & Boy\newline\textit{Xiaohang} & Learn rule-following, remember consequences, apologize, maintain good behavior \\
\bottomrule
\end{tabular}
\end{table*}

\section{APPENDIX: Distribution of Children's Coping Strategies with Representative}
Table \ref{tab:coping_strategies} presents the distribution of children's coping strategies, categorized into five primary dimensions \cite{de2009assessing}. For each dimension, the table details the specific subscales, lists the associated response codes, and provides a representative example. These examples, featuring a prompt from the \textit{TaleBot} and a child's corresponding response, serve to illustrate how each strategy was manifested in the study. These responses are coded as "Cn-m", where "n" represents participant number (1--12) and "m" indicates the story chapter (1--4).

\label{app:coping_strategies}

\begin{table}[htbp]
\centering
\caption{Distribution of Children's Coping Strategies with Representative Examples}
\label{tab:coping_strategies}
\small
\begin{tabular}{p{1.5cm}p{2.5cm}p{2.5cm}p{7cm}}
\toprule
\textbf{Dimension} & \textbf{Subscale} & \textbf{Response Code} & \textbf{Example (TaleBot Question and Child's Response)} \\
\midrule

% This section remains first, with 3 subscales.
\multirow{3}{*}{\parbox{1.5cm}{Problem Focused Coping}} 
    & Cognitive Decision Making & C3-3, C8-4, C10-4 & 
    TaleBot: \textit{"After listening to his mom, what will Xiao Ai do?"} \newline 
    C10-4: \textit{"Xiao Ai set a goal for himself. He wants to practice the things he doesn't do well a few times."} \\
    \cmidrule{2-4}
    & Direct Problem Solving & C1-2, C2-4, C3-1, C3-4, C4-1, C4-4, C5-1, C5-2, C6-1, C6-2, C6-3, C6-4, C7-1, C7-2, C8-1, C8-2, C8-3, C9-2, C9-3, C9-4, C10-1, C10-3, C11-1, C11-2, C11-4, C12-1, C12-3, C12-4 & 
    TaleBot: \textit{"Pink's little fists are clenched tightly, and her little monster is jumping around. What will Pink do next?"} \newline 
    C6-3: \textit{"Apologize to the teacher and submit the homework."} \\
    \cmidrule{2-4}
    & Seeking Understanding & C2-3, C3-2, C7-3, C10-2, C12-2 & 
    TaleBot: \textit{"Yeer sees his classmates having so much fun, but no one notices him. What would he think?"} \newline 
    C3-2: \textit{"Yeer says why didn't everyone notice me, why didn't they see me."} \\
\midrule

% This section has 2 subscales.
\multirow{2}{*}{\parbox{1.5cm}{Positive Cognitive Reframing}} 
    & Positive Thinking & C2-1, C2-2 & 
    TaleBot: \textit{"After being reminded by the teacher to listen carefully, what should Alice do?"} \newline 
    C2-2: \textit{"From now on, Alice will never feel that class is boring."} \\
    \cmidrule{2-4}
    & Optimistic Thinking & C5-4, C7-4 & 
    TaleBot: \textit{"Noting feels very sad right now, what should he do?"} \newline 
    C5-4: \textit{"The teacher will forgive him, and Noting will be happy again."} \\
\midrule

% This section has 2 subscales.
\multirow{2}{*}{\parbox{1.5cm}{Distraction Strategies}} 
    & Physical Release of Emotion & C1-1, C1-4 & 
    TaleBot: \textit{"Bunny is studying hard for the final exam, but as the exam date approaches, it starts to feel nervous and anxious. What do you think Bunny will do next?"} \newline 
    C1-1: \textit{"Take a deep breath."} \\
    \cmidrule{2-4}
    & Distracting Actions & C5-3 & 
    TaleBot: \textit{"Noting feels very sad right now, what should he do?"} \newline 
    C5-3: \textit{"Look at the blue sky outside and calm down."} \\
\midrule

% The row count for multirow is adjusted from 3 to 2, as "Wishful Thinking" is removed.
\multirow{2}{*}{\parbox{1.5cm}{Avoidance Strategies}} 
    & Avoidant Actions & C4-2, C9-1, C11-3 & 
    TaleBot: \textit{"Xiaomei sees the other children playing happily and really wants to join in, but she's a bit scared. What do you think Xiaomei will do next?"} \newline 
    C9-1: \textit{"Xiaomei stopped playing with these children and went to play with her other good friend instead."} \\
    \cmidrule{2-4}
    & Repression & C1-3 & 
    TaleBot: \textit{"Bunny remembers what his mom said and decides to do the problems he knows first. However, he is still a little worried about the problems he doesn't know. What will Bunny do next?"} \newline 
    C1-3: \textit{"Forget about it."} \\
\midrule

% This section now only has one row, so \multirow is removed.
\parbox{1.5cm}{Support Seeking Strategies}
    & Support for Feelings & C4-3 & 
    TaleBot: \textit{"Xiaoyu is standing by the window, the small stone in his heart getting heavier and heavier. What should he do?"} \newline 
    C4-3: \textit{"Go and see the mental health teacher"} \\

\bottomrule
\end{tabular}
\end{table}

\clearpage

\section{APPENDIX: School Counselors and AI Commentary Report}

\label{app:ai_analysis_report}

\begin{figure*}[h]
    \centering
    \includegraphics[width=0.8\textwidth]{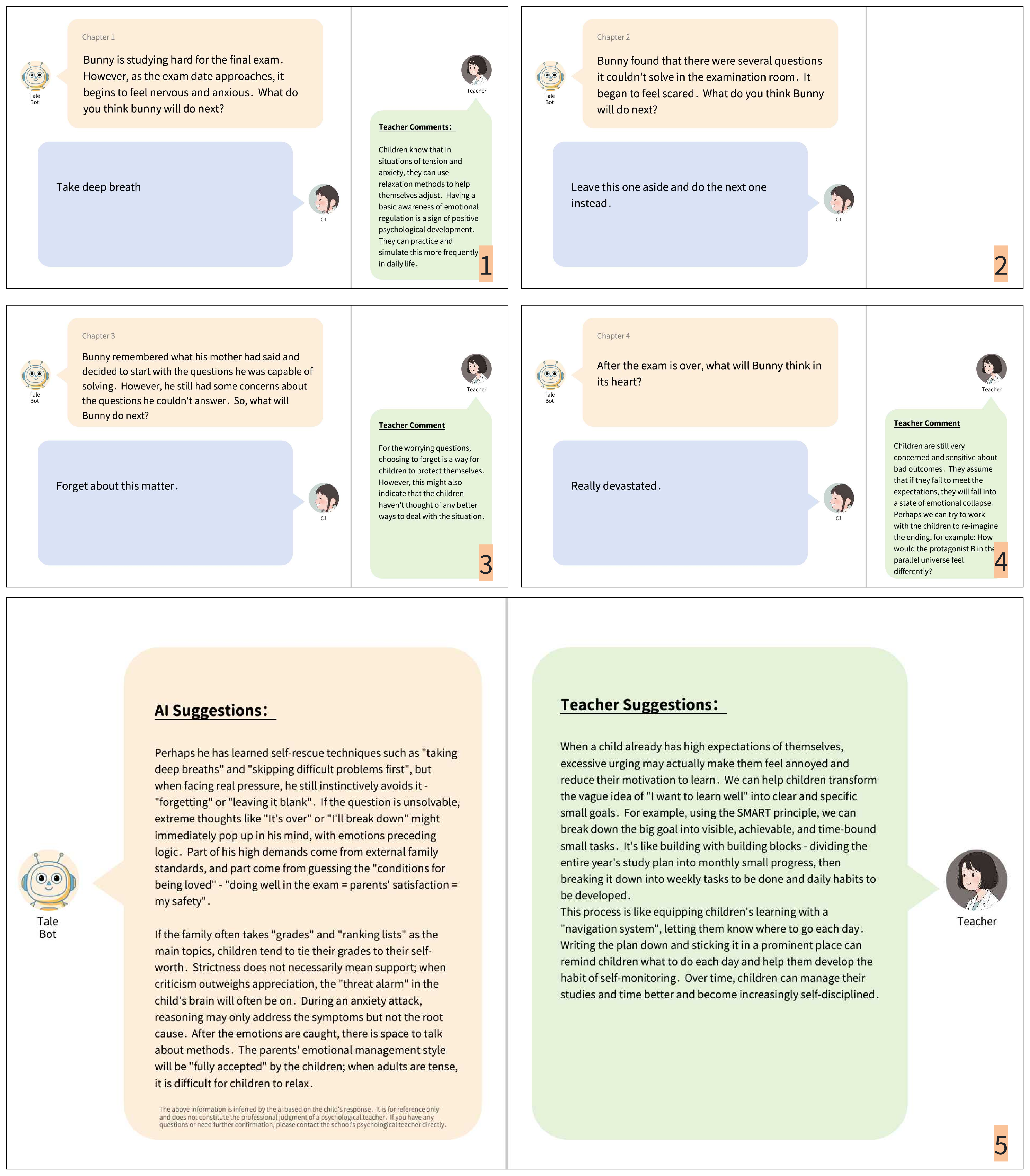}
    \caption{Annotated story book for participant C1 (translated from Chinese).}
    \label{fig:talebot_report}
\end{figure*}

Figure \ref{fig:talebot_report} shows an example of the annotated story books, the school counselor's and AI's commentaries, and advice based on the children's responses to the prompting questions.

%TC:endignore

\end{document}